%% file: main.tex
\documentclass[%
reprint,
groupedaddress,
showkeys,
preprintnumbers,
 amsmath,amssymb,
 aps,
pre,
longbibliography,
]{revtex4-2}

\newcommand{\arxiv}[1]{\href{http://arxiv.org/abs/#1}{\texttt{arXiv}:#1}}

\usepackage{tikz}
\usepackage[siunitx]{circuitikz}
\usetikzlibrary{calc}
\usetikzlibrary{circuits.ee.IEC}
\usepackage[dvipsnames]{xcolor}
\usetikzlibrary {arrows.meta} 

\usepackage{amsfonts}
\usepackage{amsmath}
\usepackage{amssymb}
\usepackage{float}

\usepackage{url}
\usepackage{amsthm}
\usepackage{graphicx}

\newcommand{\sPsi}{\left| \psi \right\rangle}
\newcommand{\sZ}{\left| 0 \right\rangle}
\newcommand{\sO}{\left| 1 \right\rangle}

\newcommand{\Complex}{\mathbb{C}}
\newcommand{\sUp}{\left| \uparrow \right\rangle}
\newcommand{\sDown}{\left| \downarrow \right\rangle}

\tikzset{ boxNode/.style= {
        rectangle,minimum size=15mm,
        very thick,draw=black
        } }

\begin{document}
\newcommand{\alert}[1]{\textbf{\textcolor{red}{#1} } }

\title{Analog Explorations of Quantum State Measurement}
\affiliation{%
Complexity Sciences Center and
Department of Astronomy and Physics\\
University of California at Davis,
One Shields Avenue, Davis, CA 95616
}%

\author{Benjamin Krawciw}

\author{James P. Crutchfield}
\affiliation{%
Complexity Sciences Center and
Department of Astronomy and Physics\\
University of California at Davis,
One Shields Avenue, Davis, CA 95616
}%

\date{\today}


\begin{abstract}
The awkward disconnect in the postulates of quantum mechanics between the
continuous, deterministic, linear, and unitary evolution governed by
Schr\"odinger's equation and the discontinuous, stochastic, nonlinear, and
nonunitary measurement process has vexed physicists for decades. Intuition from
classical mechanics, statistical mechanics, and dynamical systems theory
suggests that such a process might be modeled with a nonlinear system of
differential equations. We explore such approaches by designing a
deterministic model of an ideal projective measurement on a single qubit and
constructing an analog computer that simulates it by manipulating voltages in
an electrical circuit. The model reduces to Schr\"odinger evolution when no
measurement is taking place. During a measurement, a hidden variable that the
observer does not resolve selects the outcome, and the model is constructed so
that a uniformly distributed hidden variable yields Born-rule frequencies. We
do not derive the Born rule; rather, we show that a deterministic dynamical
system with no intrinsic noise can implement it. Over 300 trials, the circuit
reproduces the ideal model's individual outcomes to within one percent of the
decision threshold, and its outcome frequencies match the Born rule to within
about $0.02$ in probability. These results are a small but provocative step
towards modeling quantum measurement with macro-scale machines---a step of
interest for both the foundations of physics and the alternatives to digital
computing it suggests.
\end{abstract}

\keywords{quantum measurement, analog computation, Born rule, single qubit}

\preprint{\arxiv{2610.XXXXX}}

\maketitle

\newpage
\input{./intro/intro}

\input{./methods/methods}

\input{./results/results}

\input{./discussion/discussion}

\begin{acknowledgments}
We thank Fabio Anza, Jinghao Lyu, Jason R. Green, and Komal Sah for helpful
comments on an earlier draft. This material is based on work supported by, or
in part by, the Art and Science Laboratory.
\end{acknowledgments}

\input{./appendix/appendix}

\bibliography{refs}

\end{document}

%% file: intro/intro.tex
\section{Introduction}
\label{sec:intro}

We explore deterministic dynamics that bridge the gap between Schr\"odinger
time evolution and randomized state collapse during a measurement by building
an analog computer that simulates measurement on a single qubit.

It is worth stating at the outset what this does and does not show. We
construct a deterministic, nonlinear model of an ideal projective measurement
of a single qubit and build an electrical circuit that implements it. In the
model, which outcome occurs is decided by a hidden variable $\phi$ that the
observer does not resolve. The Born rule enters by design: we choose how the
outcome depends on $\phi$ so that, when $\phi$ is uniformly distributed from
the observer's point of view, the outcome frequencies are those of the Born
rule.

We therefore do not claim to derive the Born rule nor to have found the
physical mechanism behind it. Nor do we identify $\phi$ with a particular
quantum-mechanical degree of freedom. In both the model and the circuit it is
an engineered, unresolved variable. (That said, Section \ref{sec:discussion}
speculates about to what it might correspond.) What the model does demonstrate
is that a single deterministic dynamical system can reduce to Schr\"odinger
evolution when no measurement occurs and produce definite outcomes with
Born-rule statistics when one does. It does this without any intrinsic source
of noise, in contrast to the stochastic collapse theories reviewed below.

Reproducing single-qubit Born-rule statistics with a hidden variable is not
itself new \cite{Bell_1966, BohmBub1966}. With a uniformly distributed hidden
variable it is straightforward. What is new here is how the outcome comes
about. Collapse is a finite-time trajectory of a single deterministic,
noise-free system of differential equations, with no separate rule for
measurement. A measurement is simply a coupling that is switched on. And the
equations are simple enough, polynomial in the state, to build as an
electronic circuit, in which the intermediate states of a collapse are
voltages that can be recorded. We
treat only ideal projective measurements. Weak measurements and general
positive operator-valued measures (POVMs) are beyond our present scope.

There are several parallel motivations: First, we are interested in these
dynamics as a possible avenue for resolving the longstanding measurement
problem---viz., the disconnect between Schr\"odinger time evolution and
measurement. Second, the simulation of quantum problems on an analog computer
is interesting from a modern technological point of view. Even if
deterministic dynamics are ultimately unsatisfying as a resolution to the
measurement problem, it may still make for a useful and interesting method to
simulate quantum phenomena. Finally, there are pedagogical benefits. To get
started, the next section briefly reviews the measurement problem and the use of
analog computers for simulating quantum phenomena.

\section{Background}

The following first reviews basic quantum theory and quantum measurement.
It then reviews various alternative, even contemporary, views on the role of
measurement, including wave function collapse theories and the dynamical role
of decoherence.

\subsection{Postulates of Quantum Mechanics}

\newcommand{\Hilbert}{\mathcal{H}}

For our purposes, the postulates of quantum mechanics \cite[p. 166]{zet} can
be summarized as follows:
\begin{enumerate}
\item \emph{State}: A quantum system's state is given by a normalized ket vector
	$\sPsi$ in a Hilbert space $\Hilbert$. Normalized superpositions of states
	are also possible states. For a qubit $\Hilbert = \Complex^2$, so the
	state is a pair of complex amplitudes.
\item \emph{Dynamic}: Unless the system is being measured, the state evolves
	over time according to the Schr\"odinger differential equation:
\begin{align}
i \hbar \frac{d}{dt} \sPsi = \widehat{H} \sPsi,
\label{eqn:schrod}
\end{align}
	where $\widehat{H}$ is the system Hamiltonian---a linear and Hermitian
	operator. The resulting time evolution is linear and unitary.
\item \emph{Measurement}: A measurable quantity $A$ is associated with a linear
	Hermitian operator $\widehat{A}$, whose eigenvectors form an orthonormal
	basis. When $A$ is measured, the outcome is one of $\widehat{A}$'s eigenvalues $a_n$ decided at random and the state collapses to the
	corresponding normalized eigenstate $\left| v_n \right\rangle$. The
	probability of this outcome is given by the Born Rule:
\begin{align}
P (a_n) = \left| { \left\langle \psi | v_n \right\rangle } \right|^2
  ~.
\label{eqn:born}
\end{align}
	(For a degenerate eigenvalue the state is instead projected onto the
	corresponding eigenspace and renormalized. A single qubit measured in a
	basis never requires this case.)
\end{enumerate}

\subsection{The Measurement Problem}
\label{ssec:measProblem}

The measurement problem is this: The Schr\"odinger and measurement postulates
give competing rules for how a quantum state evolves over time, and there is
no clear objective criterion to decide which to use at any given moment.

It is useful to split this into two parts. The first is to say what a
measurement is: which physical processes count as measurements, and how they
can depart from Schr\"odinger evolution. The second is to say which outcome
occurs: once a measurement has taken place, what selects the single result
that the observer records? As we will see, unitary quantum mechanics with an
environment gives a good answer to the first part, but not the second.

One potential solution would be to harmonize the two postulates. Could measurement be a
special case of Schr\"odinger dynamics? The answer is no, for both closed and
open systems. The formal solution to the Schr\"odinger equation is
$\left| \psi(t) \right\rangle = \widehat{U}(t) \left| \psi(0) \right\rangle$,
where:
\begin{align}
\widehat{U}(t) = \exp \left( \frac{-i}{\hbar} \widehat{H} t \right)
\label{eqn:unitary}
\end{align}
is linear and unitary, hence invertible with inverse $\widehat{U}^\dagger(t)$.

For a closed system, measurement violates unitarity. Consider a qubit and the
two states $\left| \alpha \right\rangle = (\sZ + \sO)/\sqrt{2}$ and
$\left| \beta \right\rangle = (\sZ - \sO)/\sqrt{2}$. A measurement in the 0/1
basis may collapse either one to $\sZ$. The map is then not invertible, since
it is impossible to decide whether $\sZ$ came from $\left| \alpha
\right\rangle$ or $\left| \beta \right\rangle$. So, it cannot be unitary and
cannot be a solution of the Schr\"odinger equation.


For an open system, measurement violates linearity. Now, the system is coupled
to an environment, and basis states take the form:
\begin{align*}
\sZ \otimes \left| \epsilon_n \right\rangle \text{ or } \sO \otimes \left| \epsilon_n \right\rangle,
\end{align*}
where $\left| \epsilon_n \right\rangle$ is a basis state of the environment.
If a measurement in the 0/1 basis could be modeled by a unitary evolution
$\widehat{U}_M$, then it must leave the eigenstates $\sZ$ and $\sO$ of the
measurement alone:
\begin{align*}
\widehat{U}_M \left( \sZ \otimes \left| \epsilon \right\rangle \right) & = \sZ
\otimes \left| \epsilon' \right\rangle ~\text{and}~\\
\widehat{U}_M \left( \sO \otimes \left| \epsilon \right\rangle \right) & = \sO \otimes \left| \epsilon'' \right\rangle,
\end{align*}
where $\left| \epsilon \right\rangle$, $\left| \epsilon' \right\rangle$, and
$\left| \epsilon'' \right\rangle$ are arbitrary environment states. However,
any state can be written as
$\sPsi = \sum_n a_n \sZ \otimes \left| \epsilon_n \right\rangle + b_n \sO \otimes \left| \epsilon_n \right\rangle$,
and by the linearity of $\widehat{U}_M$:
\begin{align*}
\widehat{U}_M \sPsi = \sum_n a_n \sZ \otimes \left| \epsilon_n' \right\rangle + b_n \sO \otimes \left| \epsilon_n'' \right\rangle.
\end{align*}
Superpositions of $\sZ$ and $\sO$ therefore survive: $\widehat{U}_M$
entangles the qubit with its environment but does not collapse it. Collapse
requires a violation of linearity, so no $\widehat{U}_M$ generated by a linear
Hamiltonian can produce a single definite outcome.

This conditional unitary is nonetheless the core of the decoherence program
and of Quantum Darwinism, reviewed below. There, a measurement \emph{is} an
evolution of the form $\widehat{U}_M$: the environment changes conditioned on
the system's pointer states, here $\sZ$ and $\sO$, leaving system and
environment in a superposition of \emph{branches}, one per outcome
\cite{Touil2024}. The evolution is invertible on the joint space of system and
environment but not on the system alone. This answers the first part of the
measurement problem: it says what physical process a measurement is. It does
not answer the second. Each branch carries one outcome, but nothing in the
unitary dynamics selects which branch the observer finds.

The same linearity underlies the \emph{no-cloning theorem}
\cite{WoottersZurek1982, Dieks1982}; see also \cite[p.~532]{NielsenChuang}.
A map that copies the basis states, $\sZ \otimes \left| e \right\rangle \to
\sZ \otimes \sZ$ and $\sO \otimes \left| e \right\rangle \to \sO \otimes \sO$,
must by linearity send a superposition $(A \sZ + B \sO) \otimes \left| e
\right\rangle$ to the entangled state $A \sZ \otimes \sZ + B \sO \otimes \sO$,
not to two copies of the superposition. So no linear dynamics can copy an
unknown state. Dieks framed the result as a rebuttal to a proposed
superluminal communicator \cite{Dieks1982}: if unknown states could be cloned,
entanglement could be used to signal. A model of measurement must abandon
linearity, by the argument above, and so it also loses this automatic
protection against cloning and signaling. It must secure them some other way.
We return to this point in Sec.~\ref{sec:discussion}.

These closed- and open-system arguments motivate the following design
principles. A dynamical model of measurement must (i) reduce to Schr\"odinger evolution when
no measurement is taking place, (ii) be nonlinear and nonunitary while a
measurement is taking place, driving every state to one of the measurement's
eigenstates, and (iii) select among those eigenstates with Born-rule
frequencies.

\subsection{Decoherence and Dynamical Outcome Selection}

Despite its unitary-only approach, the decoherence program
\cite{Schlosshauer2005, Joos1985, einselection, jumps, Darwinism,
eavesdropping} in quantum physics has had remarkable success in explaining
several features of measurement through the system's coupling to its
environment. In particular, decoherence causes macroscopically-problematic
superpositions of states (such as the simultaneously dead and alive
Schr\"odinger's cat) to evolve toward unproblematic \emph{pointer} states
(either alive or dead, but not both). These appear to an observer
\cite{Joos1985, eavesdropping} as a statistical mixture of basis states. This
mechanism is called \emph{einselection} \cite{einselection}. Furthermore,
\emph{Quantum Darwinism} showed that the Born Rule can be deduced from more
modest assumptions about measurement \cite{jumps, Darwinism}. Particularly
noteworthy is the assumption that repeated measurements give the same result;
this is essentially state collapse.

Despite these successes, unitary-only quantum mechanics leaves the second part
of the measurement problem open. It lacks a complete account of collapse or,
empirically equivalently, of which branch an observer ends up in under a many
worlds interpretation \cite{Schlosshauer2005}. In a many worlds
interpretation, the wave function of the universe evolves only according to
the Schr\"odinger equation, but an observer perceives only a single
``record'', which splits into ``branches'' during the physical processes we
interpret as measurements. In any case, the time evolution seen by the
observer is necessarily nonunitary.

Resolving the second part requires something beyond unitary quantum mechanics
that selects one outcome. We group the proposals that place this selection in
the dynamics, rather than in the observer, under the name \emph{dynamical
outcome selection}. They come in three kinds: hidden-variable theories without
collapse, stochastic objective collapse theories, and deterministic collapse
theories. Our model is of the third kind.

The most prominent hidden-variable theory is \emph{pilot wave theory}
\cite{deBroglie_1928}, in which particles with definite positions are pushed
along deterministically by a wave that evolves according to the
Schr\"odinger equation without collapse. Strictly speaking, it is not a
collapse theory, since the wave never collapses. A measurement sends the
particle configuration into one branch of the wave, and the other branches,
though still present, no longer affect it. This is sometimes called
\emph{effective} collapse. Born Rule probabilities emerge as the
statistical description of ensembles of these trajectories
\cite{Valentini_2025}. The pilot wave is defined at all points in
configuration space simultaneously in a particular reference frame
\cite{Bell_1966}, so the theory is explicitly nonlocal. This is what lets it
agree with \emph{Bell's theorem} \cite{Bell_EPR}: no \emph{local}
hidden-variable theory---one in which each measurement outcome depends only on
the setting chosen at that location and on variables fixed in advance---can
reproduce all the predictions of quantum mechanics. Bell proved this by
bounding the correlations between measurements on two separated systems. In
the form usually tested \cite{CHSH}, each of two observers measures one member
of a pair along one of two settings ($a$ or $a'$, and $b$ or $b'$) and records
$\pm 1$. Writing $E(a,b)$ for the average product of the two outcomes, every
local hidden-variable theory satisfies \emph{Bell's inequality}:
\begin{align*}
S = \left| E(a,b) + E(a,b') + E(a',b) - E(a',b') \right| \le 2,
\end{align*}
whereas quantum mechanics predicts $S = 2\sqrt{2}$ for a maximally entangled
pair with suitably chosen settings. Experiments find $S > 2$---an observed
violation of Bell's inequality---including tests that close the major
loopholes \cite{HensenEtAl_BellTest1, GiustinaEtAl_BellTest2,
ShalmEtAl_BellTest3, superconductingBell}. Pilot wave theory accommodates this
violation, but at the cost of a preferred reference frame---a major paradigm
shift whose merits are still debated \cite{Valentini_2025}.

The theories in which the wavefunction does collapse are known as
\emph{objective collapse theories} \cite{Schlosshauer2005}. They maintain the
ontology of standard quantum mechanics (only states/wavefunctions), but modify the Schr\"odinger equation to
include collapse. The most popular among these are the Ghirardi-Rimini-Weber
(GRW) theory \cite{GRW}, the continuous spontaneous localization (CSL) theory
\cite{CSL}, and the Di\'osi-Penrose (DP) theory \cite{DP}. GRW describes the
wavefunction as evolving according to Schr\"odinger evolution, except with
random ``hits" that suddenly localize the wavefunction. CSL builds off GRW by
replacing the discrete hits with a continuous It\^o stochastic process. While
DP proposes spontaneous localization based on gravity. The three models all
share that they are stochastic in nature, describing a nondeterministic
evolution of the wave function, conceptually similar to Brownian motion.

Our model also modifies the Schr\"odinger equation, but it contains no noise
term. The dynamics are deterministic, and the apparent randomness of outcomes
arises entirely from the observer's ignorance of the hidden variable $\phi$.
Of the two parts of the measurement problem, our model addresses the second.
The hidden variable $\phi$ selects the outcome. We do not derive the first
part: in the model, a measurement is the switching on of a coupling to a
measurement apparatus, whose physical origin we leave open. This places it closest to the earliest deterministic collapse proposals, which
predate the stochastic theories above. Bohm and Bub \cite{BohmBub1966} added
a nonlinear term to the Schr\"odinger equation that reduces the state
continuously and deterministically to one outcome, selected by hidden
variables; averaging over those variables recovers the usual quantum
statistics. Pearle \cite{Pearle1976} likewise proposed a nonlinear term that
rapidly drives the amplitude of one component of a superposition to one and
the rest to zero. More general deterministic, nonlinear modifications of
quantum mechanics were later proposed by Weinberg \cite{Weinberg1989}, and are known to be in
tension with relativistic causality \cite{Gisin1989, Polchinski1991}. We
return to this in Sec. \ref{sec:discussion}.

\section{Deterministic Collapse on an Analog Computer}

Our goal is to explore deterministic collapse theories. Such a theory
must share most features of the Schr\"odinger equation, so that it reduces to
it in the non-measurement limit, but it must also generate the apparent
randomness and definite outcomes of the measurement postulate. The following
develops a theory for a single qubit under measurement. It then builds an
analog computer that demonstrates the theory's efficacy at re-creating apparent
randomness that arises in measurement.

We build an analog computer for two reasons: First, the analog computer acts as
a simulation of the theory, making explorations and testing easier than a
purely analytical approach. Second, the analog computer is interesting in its
own right. The approach of using one physical system to simulate another
analogous system with similar dynamics is old, older than the dominant paradigm
today of digital computers. It also underlies Feynman's suggestion for
quantum simulation---aka quantum computing \cite{Feynman1982}. Slide rules,
planimeters, and Lord Kelvin's tide predicting machine \cite{Thom81a} are
canonical examples of these analog computers. Modern research in analog
computers \cite{ThermoAI, NormalExp, Chou2019} is often motivated by time
complexity. An analog computer can sometimes sidestep lengthy calculations by
instead observing the outcome of a physical process that follows the same
dynamics.

A particularly enticing possibility is the use of analog computers as an
alternative to quantum computers. As noted, Feynman's desiderata for quantum
computing was essentially this notion of simulation of quantum phenomena by
analogous quantum physical systems \cite{Feynman1982}. Feynman even highlighted
simulating quantum systems by classical probabilistic systems as an interesting
possibility. Analog simulation of quantum computers has been explored in the
past \cite{Kish2003, Mourya_CMOS, analogQubit}. Focusing only on Schr\"odinger
evolution, none of these previous explorations attempt to model state collapse.

There have been interesting developments in the physical
simulation of pilot wave dynamics through oil droplets skittering around on a
vibrating fluid surface \cite{Bush2024}. This implementation, however, falls in
the ontology of pilot wave theory, not standard quantum mechanics.

The current roadblock to using analog computers as an effective substitute for
quantum computers is that all existing hardware implementations, including the
one we develop here, require explicit construction of the entire Hilbert space.
In contrast, a quantum computer with $N$ qubits allows operations on a $2^N$
dimensional Hilbert space (generated by the tensor product of $N$ 2D spaces)
while only requiring the physical construction of $N$ qubits. Probing this
concern, our physical implementation of the deterministic collapse model is
partly motivated by a desire to discover new ways of encoding a quantum state
in an analog system. It is conceivable that a model of state collapse could
suggest a method of encoding many dynamical variables in a small number of
physical objects, as in a quantum computer.

Our development proceeds as follows. Section \ref{sec:methods} explains the
design of our model. It discusses why the equations of motions are chosen as
they are, it summarizes the design of the electrical circuits that implement
them as an analog computer. Section \ref{ssec:analogBlocks} is a bit of a
sidebar on analog computers, included to address their likely unfamiliarity.
Then, Section \ref{sec:results} gives the results of testing this analog
computer, which confirm that the device successfully re-creates Born Rule
statistics. Finally, Section \ref{sec:discussion} summarizes our results, draws
physical implications from them, and discusses the next steps to take in the
exploration of these deterministic collapse theories.

%% file: methods/methods.tex
\newcommand\Ground{%
\mathbin{\text{\begin{tikzpicture}[circuit ee IEC,yscale=0.6,xscale=0.5]
\draw (0,2ex) to (0,0) node[ground,rotate=-90,xshift=.65ex] {};
\end{tikzpicture}}}%
}

\section{Methods}
\label{sec:methods}

This section motivates and explains the design of our analog circuit qubit.
First, Subsection \ref{ssec:ODEs} builds a system of coupled ODEs that
simulates a qubit under measurement. The design of these ODEs is constrained
to contain only the sorts of operations we can build an analog circuit to
perform. Next, Section \ref{ssec:analogBlocks} discusses analog computer
building blocks, giving an understanding of what kinds of behaviors an
analog computer built from an electronic circuit can emulate. Finally,
Subsection \ref{ssec:fullCirc} gives the full design of the analog computer in
the form of block diagrams at the high level and circuit diagrams at the
detailed level.

\subsection{Constructing Qubit ODEs}
\label{ssec:ODEs}

The equations of motion we construct are constrained by the hardware that can
solve them. Our analog computer is built from a small repertoire of standard
subcircuits, detailed in Subsection \ref{ssec:analogBlocks}A: they scale
voltages by constants, add and subtract them, integrate them over time (with a
sign inversion), and multiply pairs of them together. They do not take square
roots, exponentials, logarithms, or trigonometric functions. We therefore
restrict the right-hand sides of our ODEs to polynomials in the state
variables and parameters. This restriction drives several of the choices
below: storing the state in Cartesian coordinates rather than as an angle,
enforcing normalization with a restoring force rather than exactly, using
second-order equations, and writing the Born-rule probability as a polynomial.

The quantum mechanics reviewed in Section \ref{sec:intro} enters the model at
three points. The state, a pair of complex amplitudes, becomes a point $(x, y)$
on the unit circle (Subsection \ref{sssec:coords}). The orthonormal
eigenbasis of the measured observable becomes a pair of antipodal attractors
$\pm(c, s)$ (Subsection \ref{sssec:measForce}). And the Born rule becomes
the fraction of hidden-variable values that select each attractor
(Subsection \ref{sssec:detStoch}).

\subsection{Qubit dynamics in standard quantum mechanics}

Before building the differential equations, let us discuss the phenomena we are
trying to model. We have a single qubit, whose state takes the form:
\begin{align}
\sPsi = A \sZ + B \sO.
\label{eqn:aState}
\end{align}
For normalization, the complex coefficients $A, B \in \Complex$ must satisfy:
\begin{align}
|A|^2 + |B|^2 = 1. 
\label{eqn:norm}
\end{align}
When the qubit is not being measured, its state evolves in time according to
the Schr\"odinger Equation: 
\begin{align}
i \hbar \frac{d}{dt} \sPsi = \widehat{H} \sPsi. 
\end{align}
The Hamiltonian $\widehat{H}$ is a linear Hermitian operator, and the solution
is:
\begin{align*}
\left| \psi (t) \right\rangle = \exp \left( \frac{-i}{\hbar} \widehat{H} t \right) \left| \psi(0) \right\rangle,
\end{align*}
where the operator exponential is defined using its power series.

A general Hamiltonian may be split into linear and nonlinear parts. The
linear part alone produces unitary evolution, which cannot collapse a state
(Subsection~\ref{ssec:measProblem}), so collapse must come from the nonlinear
part; that is the part our model constructs, and it is switched on only
during a measurement. For ease of modeling we set the linear part to zero,
$\widehat{H}_\text{linear} = 0$. Little is lost by this choice. The identity
component of $\widehat{H}_\text{linear}$ only advances the global phase, which
we treat separately as a hidden variable (Sec.~\ref{sec:discussion}). Of the
remaining components, only the one proportional to $\widehat{\sigma}_y$ keeps
the amplitudes real and hence the state on the circle of
Subsection~\ref{sssec:coords}. It rotates the Bloch sphere rigidly about its
$y$ axis, which maps that circle to itself and so rotates the circle rigidly
($\dot{x} = -\Omega y$, $\dot{y} = \Omega x$). It could be added to the circuit with two
resistors cross-coupling the $x$ and $y$ integrators. Evolving under it before
a measurement is equivalent to preparing a rotated initial state, which our
experiments already sweep. The other components introduce a complex relative
phase and would require a third state variable. A linear Hamiltonian acting
\emph{during} a measurement, competing with the collapse dynamics, is left to
future work.

Thus, between measurements, $\sPsi$ is constant. In particular, the state does
not evolve toward, or settle at, any special point before a measurement: the
pre-measurement state is simply whatever state the qubit was prepared in or
left in by the previous measurement. In our experiments (Subsection
\ref{ssec:protocol}), it is the state left by the previous measurement, which
the controller records and uses to set the next measurement axis. 

When a measurement occurs, it is specified by a linear Hermitian operator
$\widehat{O}$ called an \emph{observable}. By the spectral theorem, this
operator has two orthonormal eigenstates $\sUp$ and $\sDown$. Before the
measurement, the qubit is in an arbitrary state $\sPsi$. After the measurement,
the qubit is in either $\sUp$ or $\sDown$. The probabilities of each final
state are given by the Born Rule:
\begin{align*}
\Pr (\uparrow) &= \left| { \left\langle \psi | \uparrow \right\rangle } \right|^2
~\text{and} \\
\Pr ( \downarrow ) &= \left| { \left\langle \psi | \downarrow \right\rangle } \right|^2.
\end{align*}

\subsection{Convenient coordinates for the qubit}
\label{sssec:coords}

Our challenge is to define a set of ODEs that models this measurement process
so that it can be implemented by an analog computer---that is, implemented in
classical-variable coordinates. To begin, we specify how the state is stored.
Equation (\ref{eqn:aState}) specifies the state as two complex-valued
coefficients $A$ and $B$. This requires $4$ real-valued variables to store the
same amount of information.

Removing the global phase and imposing normalization would leave the two
real parameters of the familiar Bloch sphere. We go one step further and also
fix the relative phase. This leaves $2$ real-valued variables subject to
normalization, or a single degree of freedom. First, separate the modulus and phase of the complex
variables:
\begin{align*}
\sPsi = e^{i \phi_A} |A| \sZ + e^{i \phi_B} |B| \sO. 
\end{align*}
Next, we factor out the global phase:
\begin{align*}
\sPsi = e^{i \phi_A} \left( |A| \sZ + e^{i (\phi_B - \phi_A)} |B| \sO \right).
\end{align*}
This global phase $\exp(i \phi_A)$ plays only a trivial role as an overall
constant in the Schr\"odinger Equation Eq. (\ref{eqn:schrod}) and plays no
role in the Born Rule Eq. (\ref{eqn:born}), since it is canceled by its
complex conjugate when the modulus is taken. Therefore, we can set it to $\pm
1$ without loss of generality, though it will be discussed more later.

In contrast, the remaining relative phase factor $\exp[i(\phi_B - \phi_A)]$ is
relevant physically. However, we constrain it to $\pm 1$ by restricting our
choices of observables $\widehat{O}$ and initial conditions such that all
eigenstates and initial states satisfy $\exp[i(\phi_B - \phi_A)] = \pm 1$. The
result of these simplifications is that we can store the state of the qubit as:
\begin{align*}
\sPsi = |A| \sZ \pm |B| \sO.
\end{align*}
This form can be drawn as the right half of a unit circle in the $|A|$-$|B|$
plane. 

Next, we choose to map to an entire unit circle. We do this for two reasons:
\begin{enumerate}
\item This makes orthogonal states into antipodal points. For instance, $\sZ$
maps to $(1, 0)$ and $\sO$ maps to $(-1, 0)$. This matches the up/down
intuition from spin 1/2 particles, and it also ends up making the
resulting dynamical system simpler.
\item It is easier to restrict a pair of voltages to the full unit circle than
to half the unit circle. For a full unit circle, we only need to demand $r =
1$. For a half unit circle, we would need instead $r = 1$ and $|A| \geq 0$. 
\end{enumerate}
To do this map, we use the double angle trigonometric identities to define:
\begin{align}
x & := |A|^2 - |B|^2, \\
y & := \pm 2 |A| |B|, 
\end{align}
where the sign of $y$ is chosen appropriately for the relative phase $\exp[i(\phi_B - \phi_A)] = \pm 1$.
These are components of the Bloch vector: $x = \langle \widehat{\sigma}_z
\rangle$ and $y = \langle \widehat{\sigma}_x \rangle$. (Note that our $x$
coordinate is the Bloch $z$ component.) Restricting to real amplitudes thus
confines the state to the great circle of the Bloch sphere in its $xz$ plane,
and orthogonal states are antipodal on it, just as on the Bloch sphere.

The next constraint to consider is normalization. In our simplified
coordinates, this is still:
\begin{align*}
x^2 + y^2 = 1.
\end{align*}
Imposing normalization directly would reduce the state to a single real
variable, the Bloch polar angle $\alpha$:
\begin{align}
x & = \cos(\alpha) \\
y & = \sin(\alpha), 
\label{eqn:alphaDef}
\end{align}
with $\alpha \in [0, 2\pi)$. 
However, it will prove useful to have these sines and cosines available to
intermediate stages in the model. (And, our circuits cannot compute them from
$\alpha$; see Subsection \ref{ssec:analogBlocks}.) Therefore, we continue to
express the state in terms of $x$ and $y$. We still need to impose
normalization. However, this will be done dynamically. Using multipliers and
amplifiers, we compute $r^2 = x^2 + y^2$, then design our equations to have a
forcing term that attracts $r^2$ to 1. 

\subsection{Quantum evolution ODEs without measurement}

Having settled on a coordinate system for our deterministic collapse model, we
can begin crafting the governing differential equations. The first choice to
make here is the order of our differential equations. Since the Schr\"odinger
Equation is first order, it would be intuitive to also make our model of
measurement first order. However, our integrator circuit
(Fig. \ref{fig:integrator}) inverts the output, so it is convenient to use a
pair of them instead. Additionally, having
second-order ODEs allows us to add a drag (or dissipation) term, controlled by
a parameter $d$, which smooths out the dynamics and reduces noise that may
arise in the physical circuit. With only the drag, our dynamical system so far
is:
\begin{align*}
\dot{x} & = v_x, \\
\dot{y} & = v_y, \\
\dot{v}_x & = -d v_x, \\
\dot{v}_y & = -d v_y.
\end{align*}

Now, we add in a radial force that imposes normalization to the unit circle.
This starts with a factor $(1 - r^2)$. Notice that this factor is zero when
normalization is satisfied, positive when $r$ is too small, and negative when
$r$ is too large. Next, this factor is applied to $\vec{r} = (x, y)$ to align
the forcing term with the radial direction. Lastly, this force is scaled by a
positive parameter $k$. Since normalization is meant to apply at all times, no
matter what the rest of the equation is doing, we will choose $k$ to be very
large. With the radial force included, our dynamical system is now:
\begin{align}
\label{eqn:noc1}
\dot{x} & = v_x, \\
\dot{y} & = v_y, \\
\dot{v}_x & = k x (1 - r^2) -d v_x, \\
\label{eqn:noc2}
\dot{v}_y & = k y (1 - r^2) -d v_y.
\end{align}
Equations (\ref{eqn:noc1}) - (\ref{eqn:noc2}) are sufficient to model a qubit
that is not measured. However, we will add additional forcing terms to model
the qubit while it is measured.

\subsection{Measurement}
\label{sssec:measForce}

To model measurement, we begin by setting the measurement axis, picking a point
on the unit circle with coordinates $c$ and $s$ to correspond with $\sUp$. By
our choice of coordinates, this implies $\sDown$ maps to $(-c, -s)$. Next, we
introduce a function $\theta$, whose role in measurement we define shortly.
Now, we can write out the ODEs with measurement terms as:
\begin{align}
\label{eqn:c1}
\dot{x} & = v_x, \\
\dot{y} & = v_y, \\
\label{eqn:cxdot}
\dot{v}_x & = k x (1 - r^2) -d v_x + \theta c, \\
\label{eqn:c2}
\dot{v}_y & = k y (1 - r^2) -d v_y + \theta s.
\end{align}
Via $\theta$ the new terms apply a force in the $(c, s)$ direction. If $\theta
> 0$, this will push $(x, y)$ toward $(c, s)$. If $\theta < 0$, this will push
$(x, y)$ toward $(-c, -s)$. Once $(x, y) = \pm (c, s)$, the measurement forcing
terms will compete with the radial forcing terms, but we have chosen $k$ to be
very large, so the radial terms dominate, leaving $(x, y)$ at (or very near)
$\pm(c, s)$. Now, we see that $\theta$ decides the outcome of a measurement
with its sign.

\subsection{Deterministic stochasticity}
\label{sssec:detStoch}

Quantum measurement is a key phenomenon that distinguishes the behaviors of
classical and quantum systems. Basic classical mechanics ignores measurement,
while quantum systems react substantially. To truly take on quantum dynamics
our analog qubit model must be extended to encompass the system response to
measurement. And, this entails augmenting the model to incorporate Born's rule
that maps from the quantum wave function to the probabilities of outcomes
during the measurement process.

Thus, we still must design $\theta$ to reproduce the Born Rule Eq.
(\ref{eqn:born}). Conceptually, starting with the Born Rule probability,
subtract a hidden variable $\phi$ whose value the observer does not know. For a
given $(x, y, c, s)$ configuration, $\theta$ is then positive for some range of
$\phi$ and negative for the rest. Importantly, ignorance of $\phi$ makes the
outcome appear random. Thus, the first step is to compute the Born Rule
probability for a given $(x, y, c, s)$ configuration.

To begin, express both $\langle \psi |$ and $\sUp$ in the $\sZ, \sO$ basis as
$\langle \psi | = \langle 0 | A^\dagger + \langle 1 | B^\dagger$ and $\sUp = C
\sZ + D\sO$:
\begin{align*}
P(\uparrow) & = \left| { \left\langle \psi | \uparrow \right\rangle } \right|^2 \\
& = (A^\dagger C + B^\dagger D)^2
  ~.
\end{align*}
Next, write the coefficients $A$ and $B$ in terms of an angle $\alpha$ (as
in Eq. (\ref{eqn:alphaDef})) and introduce the analogous angle
$\beta$ to rewrite $C$ and $D$:
\begin{align*}
P(\uparrow) & = \left[ \cos (\alpha / 2) \cos(\beta / 2) + \sin(\alpha / 2) \sin(\beta / 2) \right]^2
  ~.
\end{align*}
From here, manipulate the expression with the product to sum and half-angle
trigonometric identities:
\begin{align*}
P(\uparrow) & = \cos^2 \left(\frac{\alpha}{2} - \frac{\beta}{2}\right) \\
& = \frac{1 + \cos(\alpha - \beta)}{2} \\
& = \frac{1}{2} \left[ 1 + \cos(\alpha) \cos(\beta) + \sin(\alpha) \sin(\beta)
\right] ~.
\end{align*}
Last, convert the sines and cosines of $\alpha$ and $\beta$ back into $x$, $y$,
$c$, and $s$:
\begin{align}
P(\uparrow) = \frac{1}{2} \left( 1 + xc + ys \right).
\label{eqn:xyBorn}
\end{align}
Our choice to keep the $x$, $y$, $c$, and $s$ coordinates has paid off: We
compute $P(\uparrow)$ with only addition and multiplication. That is, we know
how to compute it with an analog circuit.

\begin{figure}[htbp]
\centering
\includegraphics[width=.9\columnwidth]{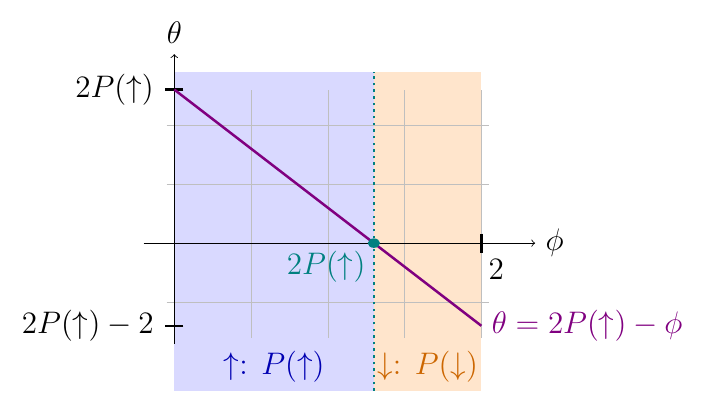}
\caption{Measurement outcome as a function of hidden variable $\phi$:
	Equation (\ref{eqn:thetaDef}) is positive (outcome $\uparrow$) for $\phi <
	2P(\uparrow)$ and negative (outcome $\downarrow$) otherwise. For $\phi$
	uniform on $[0, 2]$, the shaded (purple/yellow) fractions of the interval
	are $P(\uparrow)$ and $P(\downarrow)$, respectively.
	}
\label{fig:thetaProb}
\end{figure}

This expression identifies what the hidden variable must do. We take $\phi$ to
be a number in the interval $[0, 2]$ that is fixed for the duration of a
measurement but unknown to the observer. We assume that, lacking any
information about it, the observer assigns it a uniform probability density
over that interval:
\begin{align}
p(\phi) = \begin{cases}
1/2, & 0 \le \phi \le 2 \\
0, & \text{otherwise}
\end{cases} .
\label{eqn:phiDef}
\end{align}
This uniformity is an assumption of the model, not a consequence of it. The
construction below yields Born Rule statistics only because $\phi$ is
uniform; a nonuniform $\phi$ would yield different outcome frequencies (see
Sec. \ref{ssec:born}).
With this, we can finally construct our function $\theta$ based on Eq.
(\ref{eqn:xyBorn}) and $\phi$:
\begin{align}
\label{eqn:thetaDef}
\theta(x, y, c, s, \phi) & := 2 P(\uparrow) - \phi \\
& = 1 + xc + ys - \phi.
\end{align}
For a given $(x, y, c, s)$ configuration that yields a certain $P(\uparrow)$,
$\theta$ changes sign at $\phi = 2P(\uparrow)$: it is positive for $\phi$ in
$[0, 2P(\uparrow))$ and negative for $\phi$ in $(2P(\uparrow), 2]$, as depicted
in Fig. \ref{fig:thetaProb}. The probability that $\theta > 0$ is therefore:
\begin{align*}
\int_0^{2P(\uparrow)} p(\phi) \, d\phi = P(\uparrow) ,
\end{align*}
and the probability that $\theta < 0$ is $1 - P(\uparrow)$. In this way, the
measurement forces toward $(c, s)$ with (apparent) probability $P(\uparrow)$
and toward $(-c, -s)$ with probability $P(\downarrow) = 1 - P(\uparrow)$. With
this, we have now designed every component of the measured qubit mathematical
model.

The model itself is fully deterministic: Given $\phi$ and the initial state,
the outcome of a measurement is fixed. The randomness of the Born Rule enters
only through the observer's ignorance of $\phi$. We return to the question of
what physical degree of freedom $\phi$ might represent in Section
\ref{sec:discussion}.

\begin{figure}[htbp]
\centering
\includegraphics[width=\columnwidth]{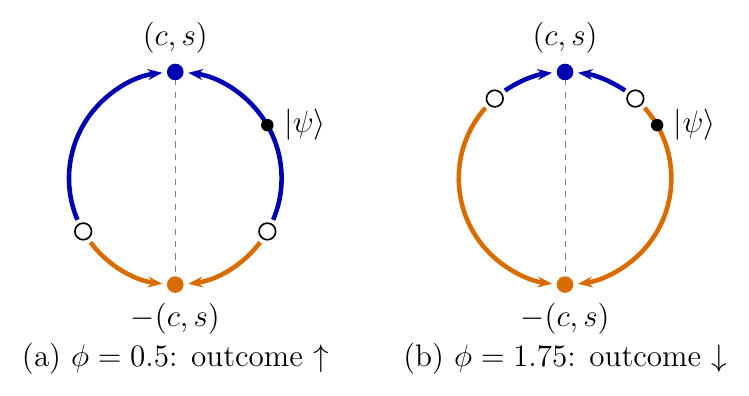}
\caption{Measurement flow on the circle of real-amplitude states, the $xz$
	great circle of the Bloch sphere, with the measurement axis $(c, s)$ at
	top. Filled dots are the attractors $\pm(c, s)$, and open dots are the
	repellers where $\theta = 0$. States in the blue basin ($\theta > 0$) flow
	to $(c, s)$, and states in the orange basin ($\theta < 0$) flow to $(-c,
	-s)$. The same prepared state $|\psi\rangle$, with $P(\uparrow) = 0.75$,
	lies in the $\uparrow$ basin for (a) $\phi = 0.5$ and in the $\downarrow$
	basin for (b) $\phi = 1.75$.
	}
\label{fig:collapseFlows}
\end{figure}

The measurement dynamics has a simple geometric picture on the circle, shown
in Fig.~\ref{fig:collapseFlows}. For fixed $\phi$, the points $\pm(c, s)$ are
attractors, and the points on the circle where $\theta = 0$ (that is, where
$xc + ys = \phi - 1$) are repellers. They split the circle into two basins.
States with $\theta > 0$ flow to $(c, s)$, and states with $\theta < 0$ flow to
$(-c, -s)$. The hidden variable moves the boundary between the basins. So, for
a given initial state, $\phi$ selects one of two flows: one in which $(c, s)$
attracts the state and $(-c, -s)$ repels it, and one with the roles reversed.

The selection holds for the whole measurement: $\theta$ cannot change sign
while the state collapses. To see this, note that the forces in Eqs.
(\ref{eqn:c1})-(\ref{eqn:c2}) are the drag plus the negative gradient of
\begin{align*}
V(x, y) = \frac{k}{4} \left( 1 - r^2 \right)^2 - \frac{\theta^2}{2},
\end{align*}
since $\nabla \theta = (c, s)$. The energy $E = \left( v_x^2 + v_y^2 \right)
/ 2 + V$ therefore satisfies $\dot{E} = -d \left( v_x^2 + v_y^2 \right) \le
0$. A measurement starts from a normalized state at rest, where $E =
-\theta_0^2/2$. Since the radial term is nonnegative, $-\theta^2/2 \le E \le
-\theta_0^2/2$ at all later times, so $|\theta| \ge |\theta_0|$. The magnitude
of $\theta$ never falls below its initial value, and its sign never changes.

\subsection{Measured Analog Qubit}
\label{sssec:modelSummary}

To summarize, the qubit model consists of:
\begin{enumerate}
     \setlength{\topsep}{-5pt}
      \setlength{\itemsep}{-5pt}
      \setlength{\parsep}{-5pt}
      \setlength{\labelwidth}{5pt}
\item A state recorded by coordinates $(x, y)$.
\item A radial force, with parameter $k$, set very high, that forces
	$x^2 + y^2 \approx 1$.
\item A drag force, parameterized by $d$. 
\item A measurement axis set by $(c, s)$ on the unit circle.
\item A measurement force that, when turned on, forces $(x, y)$ toward either
	$(c, s)$ or $(-c, -s)$, depending on the sign of a function $\theta$.
\item A function $\theta(x, y, c, s, \phi)$ that is positive with apparent
	probability $P(\uparrow)$ given by the Born Rule and negative otherwise.
\item A hidden variable $\phi \in [0, 2]$, unknown to the observer and
	therefore uniformly distributed from their point of view, which causes the
	apparent randomness. It must stay (nearly) constant while the state
	``collapses" to $\pm (c, s)$, so that $\theta$ does not change sign
	partway through a measurement.
\end{enumerate}

When the qubit is not being measured, its time evolution is given by Eqs.
(\ref{eqn:noc1})-(\ref{eqn:noc2}). When the measurement is turned on, its time
evolution is given by Eqs. (\ref{eqn:c1})-(\ref{eqn:c2}).

Beyond being implementable, the model has the structure that Section
\ref{sec:intro} asked of a dynamical model of measurement. First, with the
measurement force off it reduces to Schr\"odinger evolution: with
$\widehat{H} = 0$ the state is stationary, and the radial and drag terms
vanish for a normalized state at rest, acting only to restore normalization.
Second, with the force on, the dynamics are nonlinear and not invertible:
every initial state flows to one of two attractors, so distinct
pre-measurement states end at the same post-measurement state. Third, the
model respects the symmetries of the measurement. Since $\theta$ depends on
the state and axis only through $xc + ys$, rotating both together leaves the
dynamics unchanged, and the outcome depends only on the relative angle between
them. Relabeling the outcomes, $(c, s) \to (-c, -s)$ together with
$\phi \to 2 - \phi$, sends $\theta \to -\theta$ and leaves the measurement
force $\theta\,(c, s)$ unchanged, so neither outcome is privileged; the uniform
distribution of $\phi$ is invariant under this reflection.


\section{Electronic Analog Computers}
\label{ssec:analogBlocks}

This subsection builds a repertoire of subcircuits that carry out basic
mathematical operations, out of which an electronic analog computer can
simulate a very wide range of physical systems. This repertoire gives an
intuition for what kinds of equations are easy to simulate with an analog
computer and which are not. The circuits here are standard and found in
electronics textbooks \cite{practical, mims}.   

\subsection{Analog Computer Functional Blocks}
\label{ssec:analogCompute}

\begin{figure}[htbp]
\centering
\includegraphics[width=.9\columnwidth]{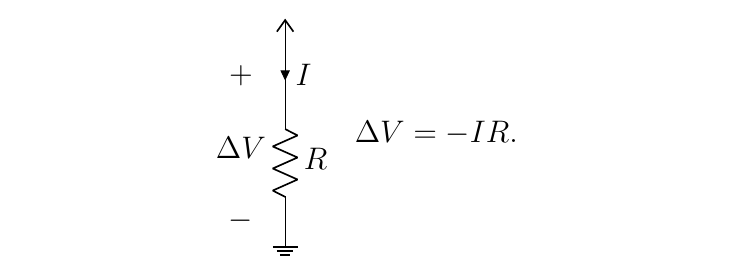}
\caption{Ohm's Law.}
\label{fig:ohm}
\end{figure}

The most basic tool at hand is Ohm's Law, which describes how to convert a
voltage to a current and scale it. If we have a resistor with resistance $R$
and we place a voltage difference $\Delta V$ across it, a current $I =
\Delta V /R$ flows from the higher voltage end to the lower voltage end.
This situation is depicted in Figure \ref{fig:ohm}. 

\begin{figure}[htbp]
\centering
\includegraphics[width=.9\columnwidth]{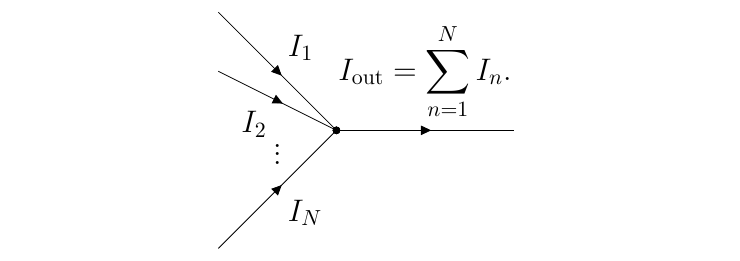}
\caption{Kirchhoff's current law.}
\label{fig:KCL}
\end{figure}

With a collection of currents formed according to Ohm's Law, we add them in
accordance to Kirchhoff's current law (KCL). KCL states that the sum of currents
into a circuit node equals the sum of currents out of the node. This is
depicted in Figure \ref{fig:KCL}.

So far, Ohm's Law and KCL restrict circuits to multiplication by positive
scalars and addition. Negative scalar multiplication and subtraction are also
needed, though. Furthermore, Ohm's Law converts voltages into currents, but
turning currents back into voltages is also important.

\begin{figure}[htbp]
\centering
\includegraphics[width=.9\columnwidth]{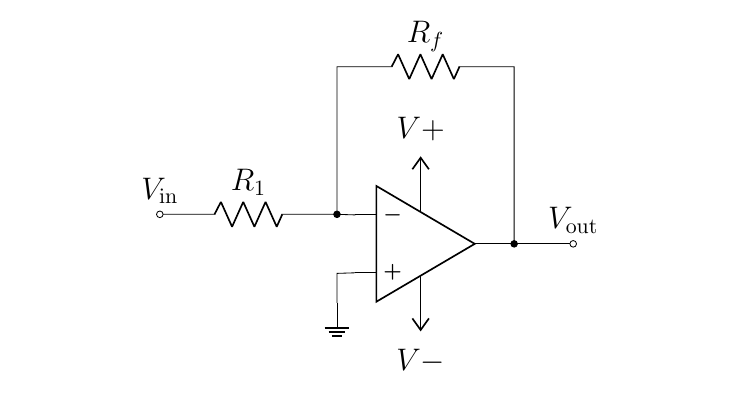}
\caption{Inverting amplifier circuit diagram: From Ohm's law, KCL, and the
	ideal op-amp assumptions, $V_\text{out} = -(R_f/R_1) V_\text{in}$.
	}
\label{fig:invAmp}
\end{figure}

All of these tasks are accomplished by the inverting amplifier (Figure
\ref{fig:invAmp}). The core of this circuit is an operational amplifier
(op-amp) \cite[p. 635-662]{practical}. With the noninverting ($+$) input tied
to ground $(\Ground)$, the amplifier equilibrates when the inverting input
($-$) is also at $0 \text{V}$ and no current flows into or out of either input.
This implies all current flowing through $R_1$ passes through the feedback resistor $R_f$, by KCL. By
Ohm's law, this becomes:
\begin{align*}
    \frac{V_\text{in} - 0\text{V}}{R_1} = \frac{0 \text{V} - V_\text{out}}{R_f}.
\end{align*}
Therefore:
\begin{align}
    V_\text{out} = -\frac{R_f}{R_1} V_\text{in}.
    \label{eqn:invAmp}
\end{align}
The output $V_\text{out}$ acts as a voltage source, supplied by the minimum and
maximum power supply voltages $V+$ and $V-$ (the voltage \emph{rails} of the
op-amp). Figure \ref{fig:invAmp} depicts a single input through $R_1$, but this
is easily extended to multiple input voltages through multiple resistors,
making use of Ohm's Law and KCL. With multiple inputs $\{V_n\}$ and associated
input resistances $\{R_n\}$:
\begin{align*}
    V_\text{out} = -R_f \sum_{n=1}^N \frac{V_n}{R_n}. 
\end{align*}

Solving differential equations requires integrating variables that are
temporal derivatives. And so, analog computers must take integrals.

\begin{figure}[htbp]
\centering
\includegraphics[width=.9\columnwidth]{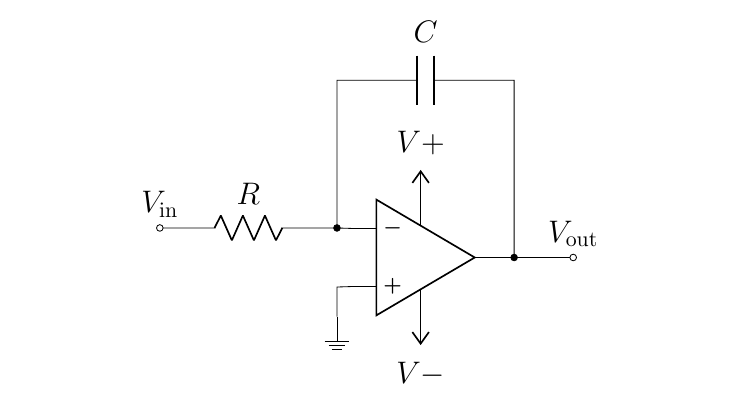}
\caption{Integrator circuit: Paralleling the logic of the inverting amplifier
	(Fig. \ref{fig:invAmp}), given that the voltage over a capacitor is
	$V_C = q/C$, $V_\text{out} = -(RC)^{-1}\int V_\text{in} dt$.
	}
\label{fig:integrator}
\end{figure}

The integrator circuit, depicted in Figure \ref{fig:integrator}, is quite
similar to the inverting amplifier, except the feedback resistor is replaced
with a capacitor. The capacitor builds up charge over time, with a
corresponding voltage across it such that $q = C V$. Since current is the flow
rate of charge, this capacitor charge is the integral of the current flowing
into it. As for the inverting amplifier, we assume that the op-amp is ideal and
equilibrated. This implies the noninverting input is held at $0 \text{V}$ and
has no current flowing into it. By KCL, all current through $R$ must flow
into the capacitor. This implies:
\begin{align*}
    \frac{V_\text{in} - 0\text{V}}{R} = \frac{d}{dt} C(0 \text{V} - V_\text{out}).
\end{align*}
Rearranging and integrating, 
\begin{align}
    V_\text{out} = -\frac{1}{RC} \int V_\text{in} dt.
    \label{eqn:integrator}
\end{align}
The product $\tau := RC$ is known as the \emph{time constant}. It has units of
seconds and sets the time scale of the integration circuit. As with the
inverting amplifier, it can be easily extended to multiple inputs on multiple
resistors. Picking a characteristic resistance $R$, this case becomes:
\begin{align*}
    V_\text{out} = -\frac{1}{RC} \int  \sum_n \frac{R}{R_n} V_n dt.
\end{align*}

With the subcircuits so far, one can simulate any linear dynamical system, such
as driven and damped simple harmonic oscillators (SHOs). However, one cannot
simulate nonlinear systems, such as the R\"ossler or Lorenz systems, which
exhibit trajectories that spread apart over time, amplifying small fluctuations
and so producing perceived randomness.

\begin{figure}[htbp]
\centering
\includegraphics[width=.9\columnwidth]{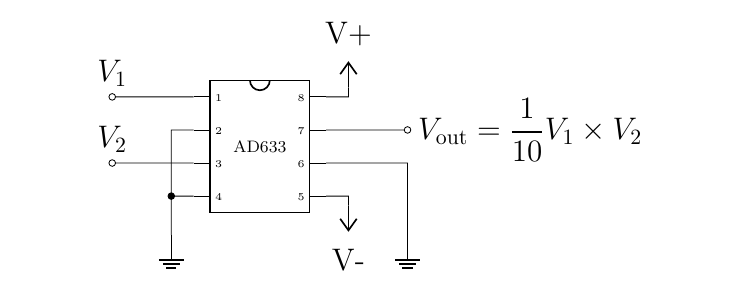}
\caption{AD633 multiplier IC pinout. For more details, see the datasheet
	\cite{ad633}.
	}
\label{fig:multiplier}
\end{figure}

To add the nonlinear terms in our nonlinear system, we need to be able to take
the product of two voltages, for example. For this, we used the prefabricated
AD633 integrated circuit (IC) \cite{ad633}. They are wired according to the
pin-out in Figure \ref{fig:multiplier}. The AD633 has two differential inputs
($V_1$ and $V_2$ on pins 1 and 3), similar to the inputs on op-amps. For
simplicity, we ground the inverting parts of the differential inputs (pins 2
and 4). For the output, there is an optional constant shift (pin 6), which is
set to zero by grounding. The IC is supplied with external power via positive
($V^+$, pin 8) and negative voltage ($V^-$, pin 5) rails (also like an op-amp).
The output, the product $V_\text{out} = V_1 \times V_2 / 10$, appears on pin 7
scaled down by a factor of 10.

To review, we now have circuits that allow us to convert between currents and
voltages, scale by positive real numbers, add currents, invert currents and
convert them to voltages, take integrals, and take the product of two voltages.
This set of operations is enough to simulate nontrivial ODEs, but there are
still some operations that are off limits. In particular, we have not discussed
circuits that take square roots, exponentials, logarithms, or trigonometric
functions. We could, hypothetically, approximate these with a Taylor series,
using stacked AD633 multipliers for higher order terms, but this would be
unnecessarily resource intensive.

There are, however, analog circuits that perform logarithms and exponentials.
For example, replacing the integration capacitor with a diode in the feedback
loop gives a logarithmic operation, as the diode current depends exponentially
on the voltage. Similarly, replacing the input resistor with a diode gives an
anti-logarithmic operation. Using these provides a much wider range of ODEs that
can be simulated. These possibilities aside, we treated the list of operations
above as exhaustive; this is the design constraint that shaped the qubit model
of Subsection \ref{ssec:ODEs}.

\subsection{Analog Implementation}
\label{ssec:fullCirc}

\begin{figure}[htbp]
\includegraphics[width=.9\columnwidth]{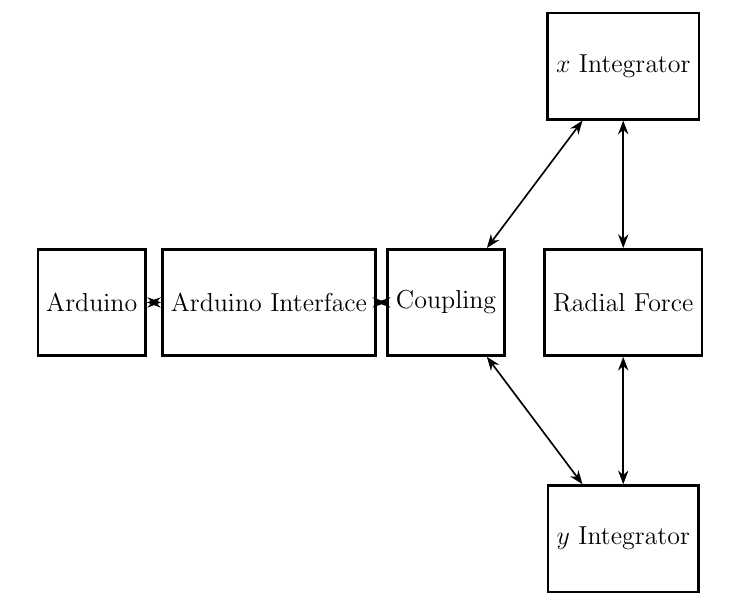}
\caption{Block diagram for the analog computer implementation of a qubit.}
\label{fig:bigblock}
\end{figure}

With the mathematical model set and the functional blocks in hand, we can
explain the design of the analog computer that implements the qubit model. A
block diagram for the full computer is given in Fig. \ref{fig:bigblock}. The
major components are:
\begin{enumerate}
     \setlength{\topsep}{-5pt}
      \setlength{\itemsep}{-5pt}
      \setlength{\parsep}{-5pt}
      \setlength{\labelwidth}{5pt}
\item Double integrators for $x$ and $y$: Perform the integrals that solve the
	differential equations and store the current state coordinates.
\item Radial force module: Computes $1 - r^2$ and returns that to the
	$x$ and $y$ integrators.
\item Microcontroller: Initially, sets the $(c, s)$ coordinates of the
	measurement axis and the hidden variable $\phi$, turns the measurement
	force on and off (off corresponds to $c = s = 0$), and records data.
\item Interface module: Uses amplifier circuits to shift between the voltages
	of the microcontroller and the voltages used by it.
\item Coupling module: Computes $\theta$ and then subsequently the measurement 
	forcing terms. 
\end{enumerate}

Not shown in the block diagrams is the power supply. All op-amps in the
circuit are LF411s \cite{lf411}, and they and the AD633 multipliers run from
$\pm 15$\,V rails provided by a benchtop supply (labeled G in
Fig. \ref{fig:circuitPhoto}).

\begin{figure}[htbp]
\includegraphics[width=.9\columnwidth]{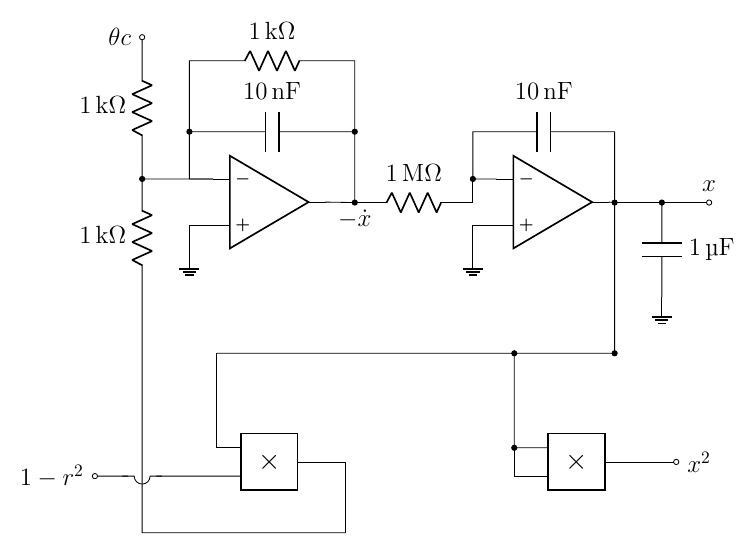}
\caption{Double integrator circuit: Computes $x$ from currents corresponding to
	$\dot{v}_x$ by integrating twice. The first (damped) integrator outputs
	$-\dot{x}$; the second outputs $x$. The $y$ circuit is identical, with
	$\theta s$, $-\dot{y}$, $y$, and $y^2$ in place of $\theta c$, $-\dot{x}$,
	$x$, and $x^2$. Boxes marked $\times$ are AD633 multipliers
	(Fig.~\ref{fig:multiplier}), each outputting $V_1 V_2 / 10\,\mathrm{V}$;
	their power connections are omitted here and in Figs.~\ref{fig:theta}
	and \ref{fig:thetacs}.
	}
\label{fig:xy}
\end{figure}

The first major subcomponents are the double integrators for $x$ and $y$. As
the name suggests, these take currents corresponding to the terms in
$\dot{v}_{x/y}$ Eq. (\ref{eqn:cxdot}) and integrates them twice to compute
$x$ or $y$. The circuit diagram for $x$ is shown in Fig. \ref{fig:xy}. The
circuit for $y$ is identical, except it takes input from $\theta s$ instead of
$\theta c$ and outputs $y$ and $y^2$ instead of $x$ and $x^2$.

\begin{figure}[htbp]
\includegraphics[width=.9\columnwidth]{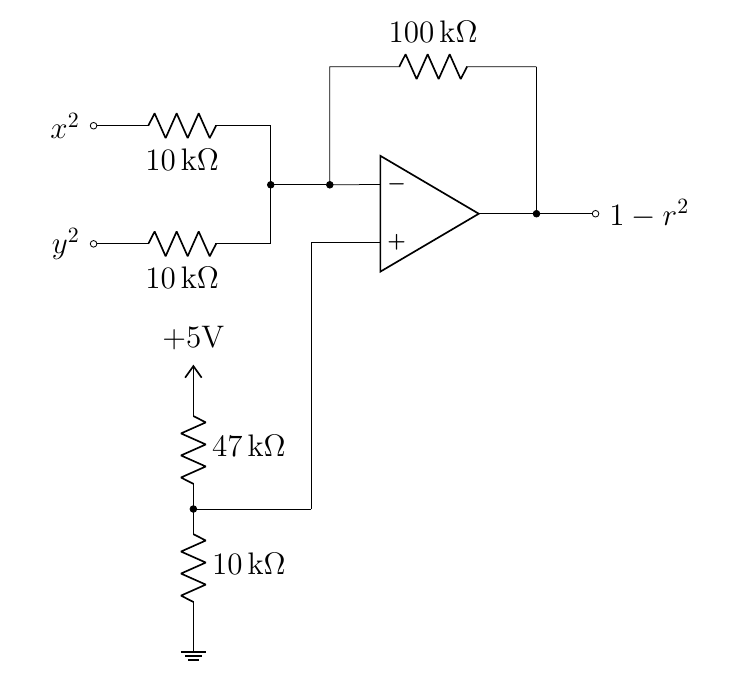}
\caption{Radial force module: Compute the displacement of $(x, y)$ from the
	unit circle, which is taken as input to the $x/y$ integrators to create a
	restoring force that keeps the state on the unit circle.
	}
\label{fig:rsquare}
\end{figure}

Next is the radial force module, which computes $1-r^2$ to determine if $x$ and
$y$ are properly positioned on the unit circle. This is depicted in Fig.
\ref{fig:rsquare}. The inputs are the multiplier outputs $V_x^2/10\,\mathrm{V}$
and $V_y^2/10\,\mathrm{V}$, and the constant term comes from a voltage divider
on the $+5$\,V rail feeding the noninverting input. With a $47\,\mathrm{k}\Omega$
over $10\,\mathrm{k}\Omega$ divider and a noninverting gain of 21, the module
outputs
\begin{align*}
V_\text{out} = 25\,\mathrm{V} \left[ \frac{14}{19}
	- \left( \frac{V_x}{5\,\mathrm{V}} \right)^2
	- \left( \frac{V_y}{5\,\mathrm{V}} \right)^2 \right] ,
\end{align*}
so the state settles on a circle of radius $R_\text{eq} = \sqrt{14/19} \approx
0.858$ natural units ($\approx 4.3$\,V) rather than the unit circle ($5$\,V).
This is a deliberate compromise. A radius of exactly $5$\,V would require a
divider ratio of $3.2$ (e.g., $32\,\mathrm{k}\Omega$ over $10\,\mathrm{k}\Omega$),
which is not a standard off-the-shelf resistor value, so the circuit was
built from standard resistors and the resulting radius is compensated in how
$\phi$ is sampled (Subsection \ref{sssec:scaling}).

\begin{figure}[htbp]
\includegraphics[width=.9\columnwidth]{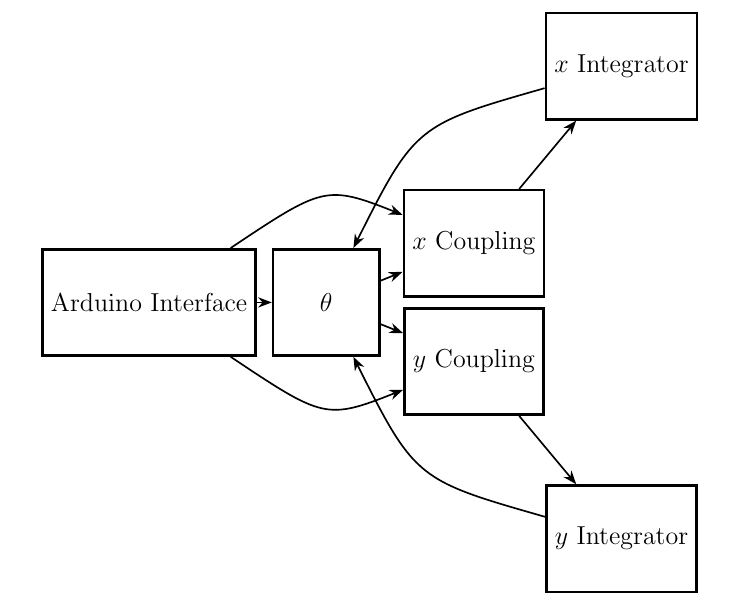}
\caption{Coupling module: Computes measurement forces on the qubit.}
\label{fig:cblock}
\end{figure}

The coupling module is complex enough to warrant its own block diagram,
depicted in Fig. \ref{fig:cblock}.

\begin{figure}[htbp]
\includegraphics[width=.9\columnwidth]{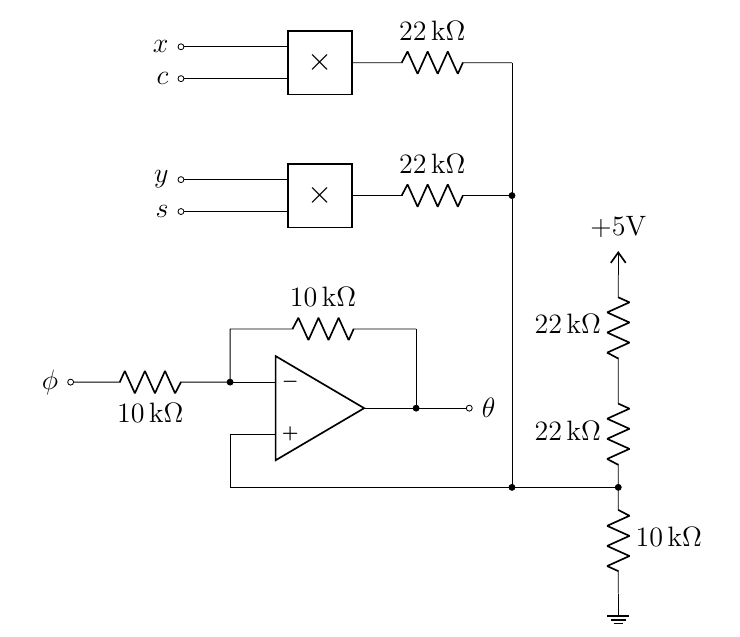}
\caption{Subcircuit that computes the $\theta$ term of the qubit model, which 
	decides measurement outcomes.
	}
\label{fig:theta}
\end{figure}

As discussed in Subsection \ref{ssec:ODEs}, $\theta$ is the term that decides
which measurement outcome is chosen, in accordance with the Born Rule with the
hidden variable $\phi$ as the source of apparent randomness. The circuit
computing $\theta$ is documented in Fig. \ref{fig:theta}.

\begin{figure}[htbp]
\includegraphics[width=.9\columnwidth]{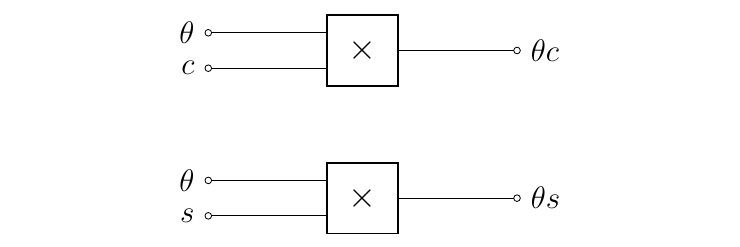}
\caption{Subcircuit that computes the measurement force terms $\theta c$ and
	$\theta s$.
	}
\label{fig:thetacs}
\end{figure}

The next subcircuits of the coupling module are the $x/y$ coupling terms. These
take $\theta$ and the measurement axis coordinates $(c, s)$ as input,
multiplies them, and computes the forcing terms $\theta c$ and $\theta s$ that
are fed into the integrators to perform measurement on the qubit. These
subcircuits are diagrammed in Fig. \ref{fig:thetacs}.

The last module is the Arduino microcontroller interface module, consisting of
several amplifiers that convert between Arduino voltages (ranging from $0$V to
$5$V) and the voltages used in the analog computer (from $-5$V to $5$V). For
$c$, $s$, and $\phi$, these amplifiers draw from the PWM (pulse-width
modulation) pins on the microcontroller, which mimic analog voltages by
rapidly switching a digital signal on and off. To smooth these out,
each PWM output feeds a second-order active low-pass filter in the Sallen-Key
topology \cite{sallenKey}, which places two resistor-capacitor stages around an
op-amp. A low-pass filter passes slowly varying signals (here, the intended
steady voltage) and suppresses rapidly varying ones (the PWM switching). The
output of a second-order filter obeys the same equation as a driven, damped
harmonic oscillator:
\begin{align}
\ddot{V}_\text{out} + \frac{\omega_0}{Q} \dot{V}_\text{out}
	+ \omega_0^2 V_\text{out} = K \omega_0^2 V_\text{in} ,
\label{eqn:filterODE}
\end{align}
so the filter is characterized by three numbers. The gain $K$ is the ratio
$V_\text{out}/V_\text{in}$ for a constant input. The natural frequency
$\omega_0$ sets where suppression begins: inputs oscillating well below
$\omega_0$ pass through (multiplied by $K$), while the amplitude of inputs
well above $\omega_0$ falls off as $(\omega_0/\omega)^2$. The quality factor
$Q$ is the usual measure of an oscillator's damping: $Q = 1/2$ is critically
damped, and larger $Q$ is underdamped, producing a resonant peak near
$\omega_0$. Both filters are designed for a Butterworth response, $Q =
1/\sqrt{2}$, which keeps the gain as flat as possible below $\omega_0$ without
a resonant peak \cite{vanValkenburg}. For this choice, $\omega_0 = 2\pi
f_{-3\,\mathrm{dB}}$, the frequency at which the output amplitude has fallen to
$1/\sqrt{2}$ of its low-frequency value. We quote suppression in decibels (dB),
$20 \log_{10}$ of an amplitude ratio, so that each factor of 10 in frequency
above $\omega_0$ adds $40$\,dB of suppression.

\begin{figure}[htbp]
\includegraphics[width=.9\columnwidth]{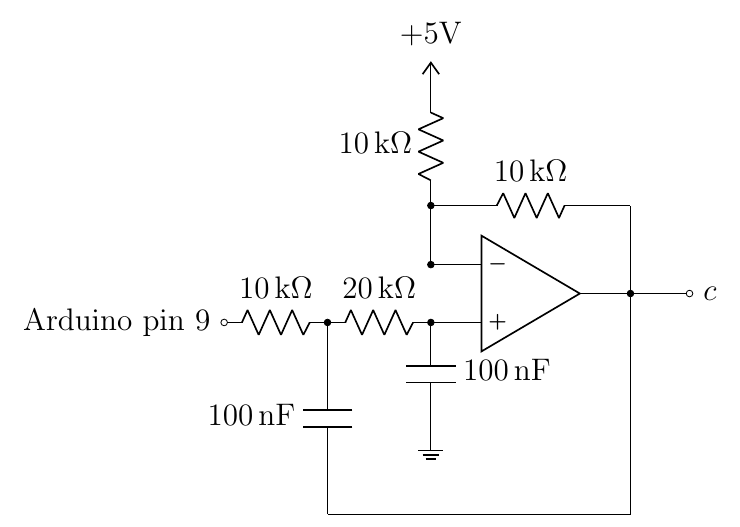}
\caption{Circuit that converts a PWM signal from the microcontroller (pin 9)
	into a stable voltage for $c$. The gain-2, $+5$V-referenced level
	shifter doubles as the amplifier of a second-order Sallen-Key low-pass
	filter. With equal capacitors and $K=2$, $Q=\sqrt{R_1/R_2}$, so
	$R_2 = 2R_1$ gives a Butterworth response with $f_{-3\,\mathrm{dB}}
	\approx 113$\,Hz, attenuating the 31.4\,kHz PWM carrier by
	$\approx 98$\,dB. An identical circuit on pin 10 drives $s$.
	}
\label{fig:ard1}
\end{figure}

\begin{figure}[htbp]
\includegraphics[width=.9\columnwidth]{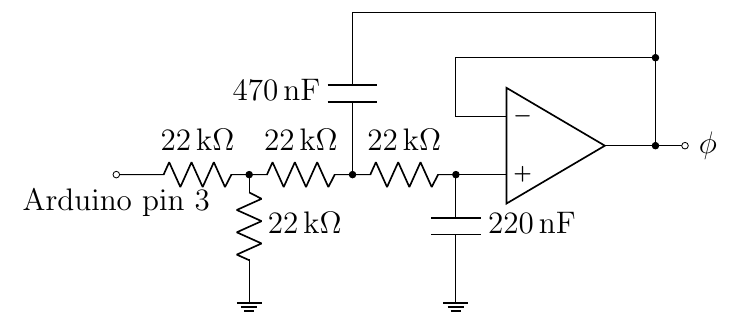}
\caption{Circuit that converts a PWM signal from the microcontroller (pin 3)
	into a stable voltage for $\phi$. A 2:1 input divider (DC gain $1/2$,
	leaving headroom at the top of the PWM range) feeds a unity-gain
	second-order Sallen-Key low-pass filter. Including the divider's
	Th\'evenin resistance, $Q \approx 0.72$ (near Butterworth) and
	$f_{-3\,\mathrm{dB}} \approx 18$\,Hz, attenuating the $\approx 490$\,Hz
	PWM carrier by $\approx 57$\,dB.
	}
\label{fig:ardPhi}
\end{figure}

\begin{figure}[htbp]
\includegraphics[width=.9\columnwidth]{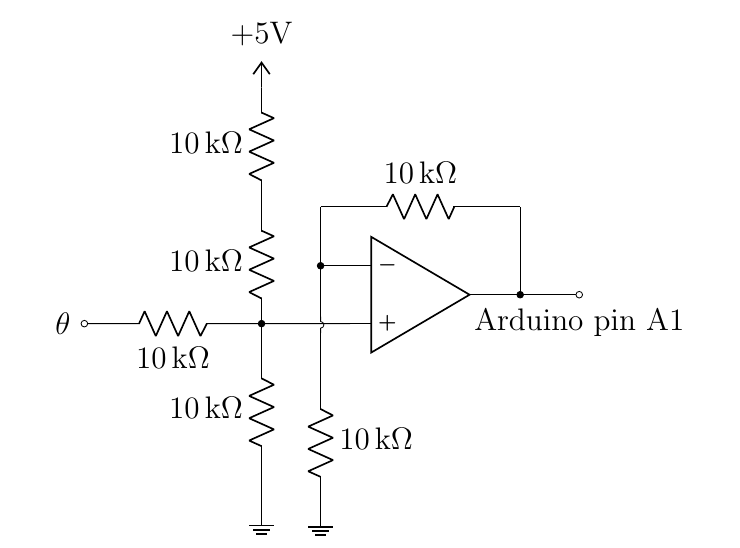}
\caption{Level-shifting amplifier for $\theta$, allowing $\theta$ to be read
	and recorded by the microcontroller (analog pin A1). The pin voltage is
	$2\,\mathrm{V} + 0.8\,V_\theta$.
	}
\label{fig:ardTheta}
\end{figure}

\begin{figure}[htbp]
\includegraphics[width=.9\columnwidth]{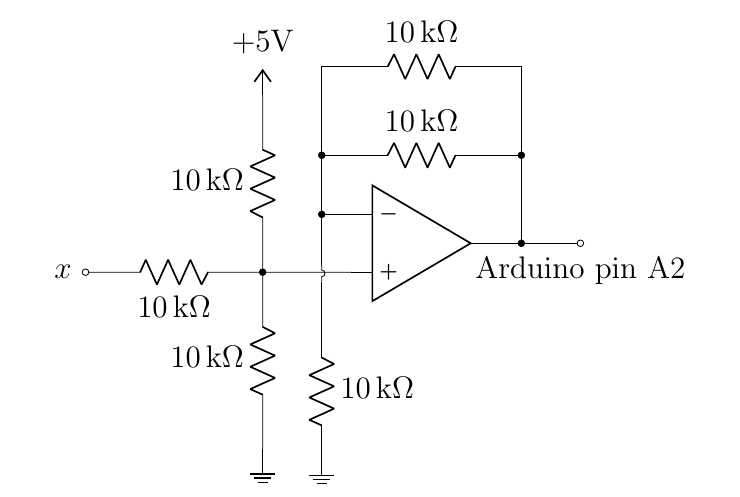}
\caption{Level-shifting amplifier for $x$, allowing it to be read and recorded
	by the microcontroller (analog pin A2). The pin voltage is
	$2.5\,\mathrm{V} + 0.5\,V_x$. A copy of this amplifier on pin A3 reads $y$.
	}
\label{fig:ard2}
\end{figure}

For the other amplifiers, the
general process starts with analog voltages in the model, amplifies them to
map nicely onto the $0-5$V range, then feeds the outputs to the
microcontroller's analog pins for measurement and recording. These are
depicted in Figs. \ref{fig:ard1} - \ref{fig:ard2}. 

\subsection{Voltage scaling and calibration}
\label{sssec:scaling}

The circuit represents the model variables $x$, $y$, $c$, and $s$ with
$5$\,V per natural unit, while $\theta$ and $\phi$ have their own scale of
$50/47$\,V per natural unit, set by the resistor ratios in the coupling
module. The one departure from the ideal model, the equilibrium radius set by
the radial force module, is compensated in how $\phi$ is chosen rather than in
hardware.

As described above, the radial force module (Fig. \ref{fig:rsquare}) holds
the state on a circle of radius $R_\text{eq} = \sqrt{14/19} \approx 0.858$
instead of the unit circle. The measurement axis $(c, s)$ is
still set to unit length, so $xc + ys = R_\text{eq} \cos \Delta$, where
$\Delta$ is the angle between the state and the measurement axis. The
$\theta$ circuit (Fig. \ref{fig:theta}), however, was designed for $r = 1$ and
computes:
\begin{align*}
V_\theta = \frac{50}{47}\,\mathrm{V} \left( 1 + xc + ys \right) - V_\phi ,
\end{align*}
so $\theta$ changes sign at $V_\phi = (50/47)(1 + R_\text{eq} \cos \Delta)$\,V.
Sampling $V_\phi$ uniformly over $[0, 100/47]$\,V, the image of $\phi \in [0, 2]$,
would therefore compress the probabilities toward $1/2$. Instead, $V_\phi$ is
sampled uniformly over
\begin{align*}
V_\phi \in \frac{50}{47} \left[ 1 - R_\text{eq},\, 1 + R_\text{eq} \right] \mathrm{V}
\approx [0.151, 1.977]\,\mathrm{V} ,
\end{align*}
for which the fraction of $\phi$ values with $\theta > 0$ is
$(1 + \cos \Delta)/2$, exactly the Born Rule probability
Eq. (\ref{eqn:xyBorn}) for a normalized state. These endpoints come from the
circuit analysis, not from a fit to data.

The microcontroller is an Arduino (ATmega328P). The $c$ and $s$ drives use
pins 9 and 10 with Timer1 reconfigured for $31.4$\,kHz PWM; a duty cycle $D$
gives $V_c = (2D - 1) \times 5$\,V after the filter of Fig. \ref{fig:ard1}.
The measurement force is switched off by setting $c = s = 0$; because an
8-bit duty cycle cannot land exactly on $50\%$, the nearest code leaves
$V_c = V_s \approx +20$\,mV ($c = s \approx 0.004$), half of one PWM step. The $\phi$
drive uses pin 3 at the default $490$\,Hz; after the 2:1 divider of
Fig. \ref{fig:ardPhi}, $V_\phi$ spans $0$ to $2.5$\,V in steps of about
$9.8$\,mV, giving roughly 186 distinct levels across the sampling interval.
The logged value of $\phi$ is the quantized value actually delivered.
The microcontroller holds $\phi$ fixed during each trial. Because we choose
its value rather than leaving it unknown, we can both sample the uniform
distribution of Eq. (\ref{eqn:phiDef}) efficiently (with a stratified grid)
and compare each recorded outcome with the one the ideal model predicts for
the same $\phi$. The sampling protocol is described in Section
\ref{sec:results}.
The voltages $\theta$, $x$, and $y$ are read back through the level shifters of
Figs. \ref{fig:ardTheta} and \ref{fig:ard2} by the microcontroller's 10-bit
analog-to-digital converter.

%% file: results/results.tex
\section{Results}
\label{sec:results}

\begin{figure}[htbp]
\centering
\includegraphics[width = \columnwidth]{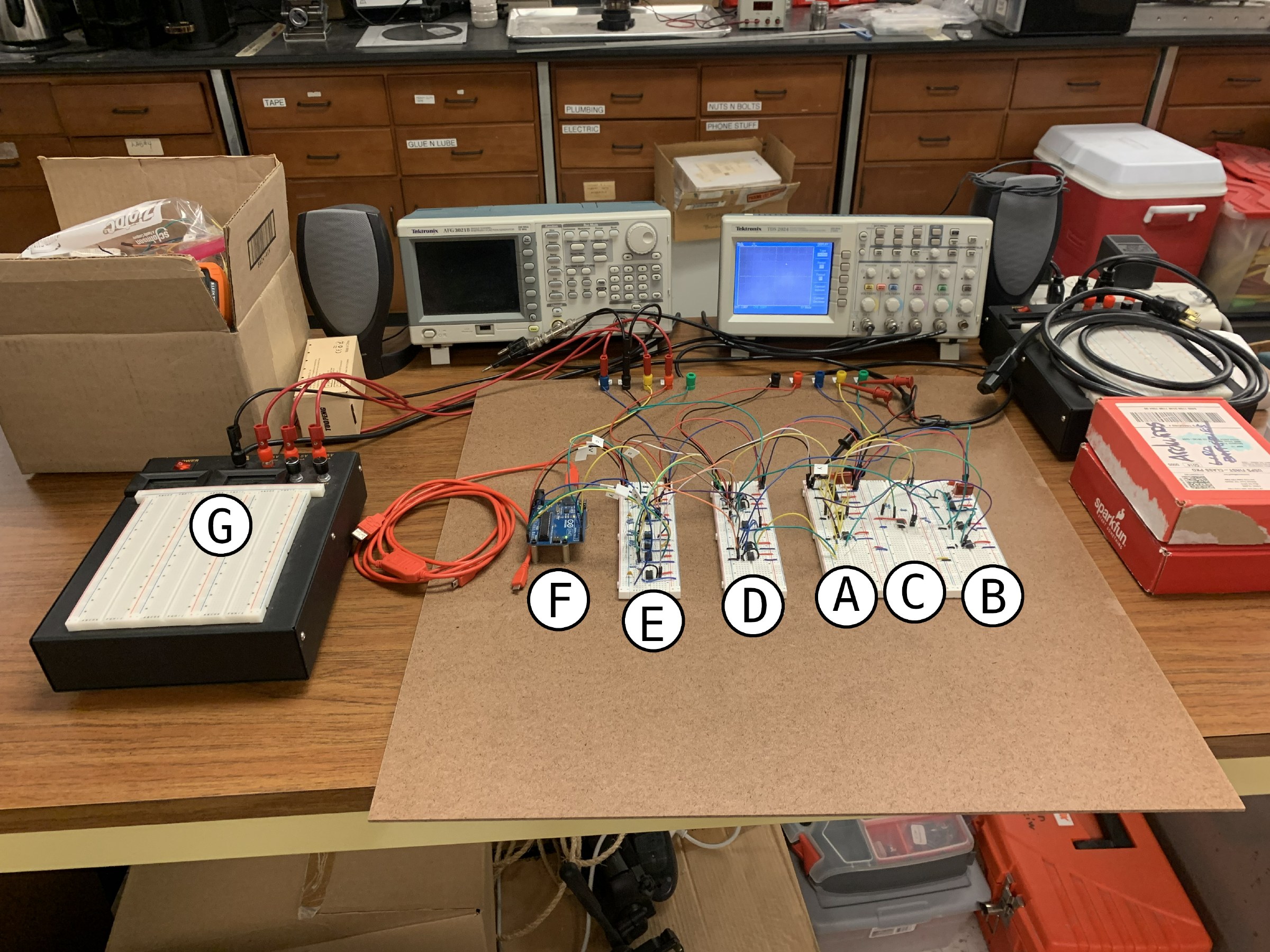}
\caption{Measured qubit analog computer circuit built according to Section
	\ref{ssec:fullCirc}'s design. Major subcircuits: (A) $x$ Integrator, (B)
	$y$ Integrator, (C) Radial Force, (D) Coupling, (E) Arduino Interface, (F)
	Arduino, and (G) $\pm 15$\,V power supply. Test equipment in the background
	is not part of the circuit; the oscilloscope is used only for monitoring.
	}
\label{fig:circuitPhoto}
\end{figure}

We built the analog computer designed in Subsection \ref{ssec:fullCirc}. A
picture of the computer is given in Fig. \ref{fig:circuitPhoto}. The Arduino
microcontroller conducts a test suite and records data that answer two
questions. First, does the circuit actually implement the ODE model of
Subsection \ref{ssec:ODEs}, to within the tolerances of its components?
Second, do the empirical probabilities of the measurement outcomes obey the
Born Rule?

To answer them, we perform an array of trials. In each trial the
controller sets a measurement axis at a predetermined angle away from the
model's current state and holds the hidden variable $\phi$ at a chosen value,
then turns the measurement force on, lets the model equilibrate at a new
state, and records where it settled. Varying both the relative angle and
$\phi$ across trials, we compare the observed statistics to the Born Rule.
Since the circuit is deterministic, we can also compare every individual
outcome to the one the ideal model predicts for the same angle and $\phi$.

\subsection{Test protocol}
\label{ssec:protocol}

\begin{figure}[htbp]
\centering
\includegraphics[width = \columnwidth]{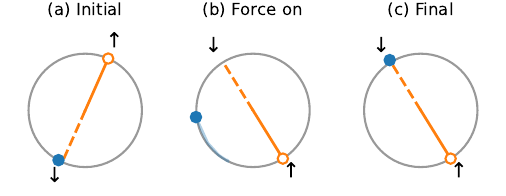}
\caption{Snapshots of trial 46 (relative angle $\Delta = 60^\circ$, so
	$P(\uparrow) = 0.75$), in natural units: (a) The force is off and the state
	(blue dot) rests at the pole where the previous measurement left it, shown
	with that measurement's axis (orange; solid toward $\uparrow$, dashed
	toward $\downarrow$). (b) A new axis is set $60^\circ$ away, the force is
	switched on, and the state begins moving; the snapshot is taken $0.11$\,s
	later, with the pale trail showing the path so far. (c) About one second
	later the state has equilibrated at $\downarrow$. For this value of
	$\phi$ the ideal model also predicts $\downarrow$.
	}
\label{fig:exampleMmt}
\end{figure}

\begin{figure}[htbp]
\centering
\includegraphics[width = \columnwidth]{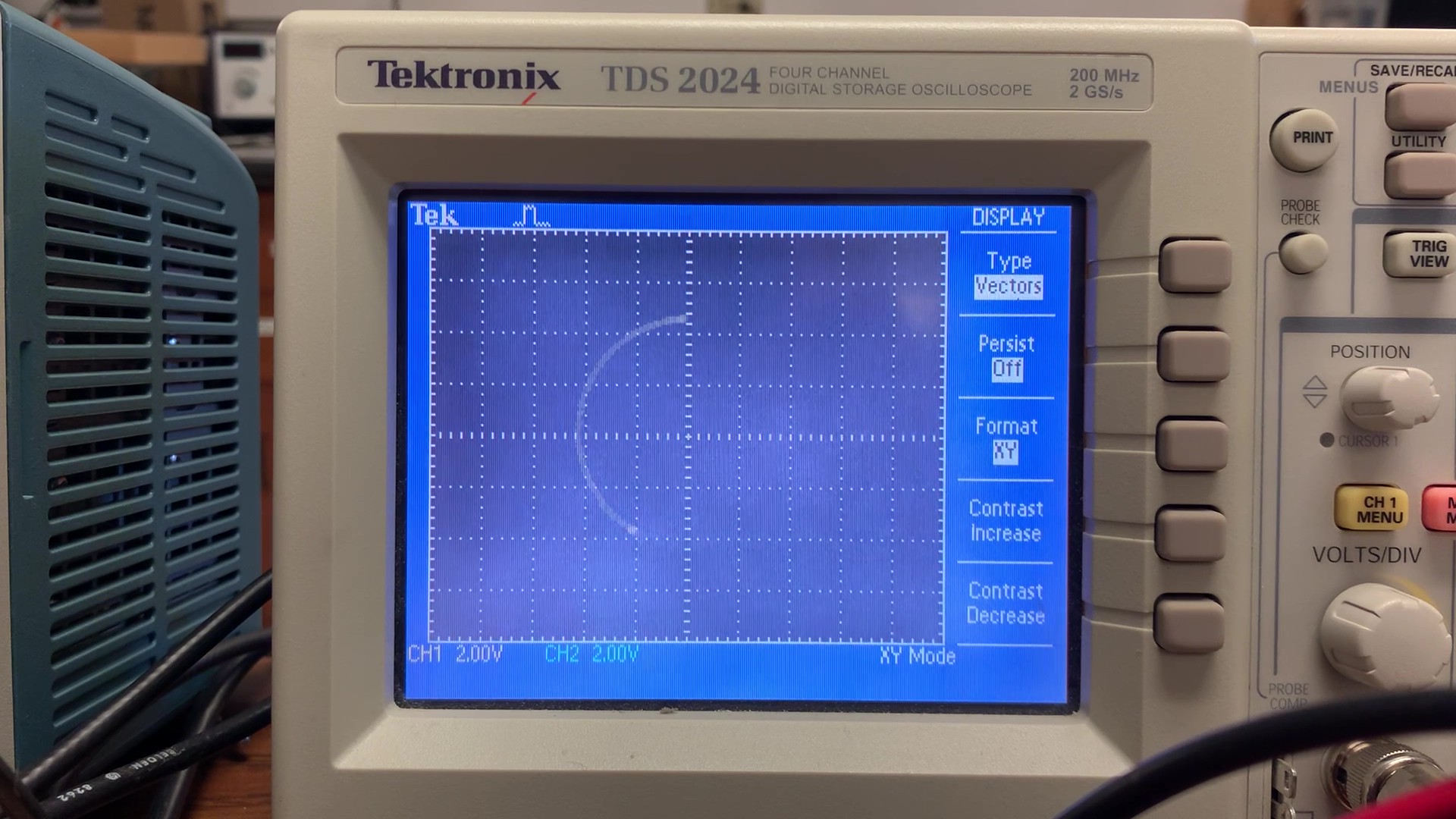}
\caption{Oscilloscope trace during a trial: The oscilloscope plots the $(x, y)$
	voltages at a given time. The trace persists for a time when the voltage
	changes, leaving a visible trail. The brightest yellow dot is the current
	voltage, while the fading tail depicts where the voltage was previously. In
	this case, we can see $(x, y)$ is rapidly collapsing to one of the
	measurement equilibria.
	}
\label{fig:oscFrame}
\end{figure}

\begin{figure}[htbp]
\centering
\includegraphics[width = \columnwidth]{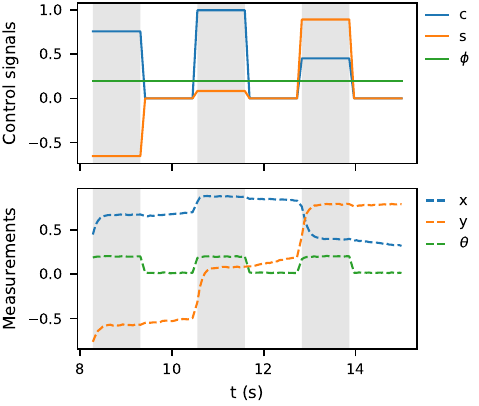}
\caption{Signals recorded during three consecutive trials (trials 3--5), with
	the force-on windows shaded: (top) The control signals. When the force is
	on, $(c, s)$ is set at the trial's relative angle from the state; when it
	is off, both are set to zero, removing the measurement terms from the
	ODEs. The hidden variable $\phi$ is held constant, since all three trials
	share one of the 30 values of $\phi$. (bottom) The recorded state. $\theta$
	decides the outcome: if $\theta > 0$, $(x, y)$ is driven toward $(c, s)$,
	and if $\theta < 0$, toward $(-c, -s)$. Here $\theta > 0$ in all three
	trials and $(x, y)$ settles near $(c, s)$ each time. The slow change in
	$(x, y)$ during the force-off windows is the drift discussed in Subsection
	\ref{ssec:limitations}.
	}
\label{fig:trialCycle}
\end{figure}

Each trial begins with the measurement force switched on. Just before this,
the controller takes the last recorded $(x, y)$, the final reading of the
preceding force-off window, and sets the new measurement axis $(c, s)$ at the
trial's relative angle $\Delta$ from it. The force stays on
for about one second while $(x, y)$ collapses to one of the two poles
$\pm(c, s)$. The force is then switched off (by setting $c = s = 0$) for about
one second, during which the radial force alone holds the state on the circle
and the state should remain at rest. That resting state is the starting point
of the next trial. This process is depicted in Fig. \ref{fig:exampleMmt}.
Figure \ref{fig:oscFrame} shows an example of what is seen on the oscilloscope
monitor while the model runs, and Fig. \ref{fig:trialCycle} plots the relevant
signals during three consecutive trials.

The particular parameters used in testing are:
\begin{itemize}
     \setlength{\topsep}{-5pt}
      \setlength{\itemsep}{-5pt}
      \setlength{\parsep}{-5pt}
      \setlength{\labelwidth}{5pt}
\item The hidden variable takes $N = 30$ values on a midpoint grid: the
	sampling interval of Subsection \ref{sssec:scaling} is divided into 30
	equal cells and $\phi$ is set to the center of each. A midpoint grid is
	used because the fraction of its points below any threshold is an unbiased
	estimate of the corresponding probability; a grid that includes the
	endpoints biases the estimate toward $1/2$.
\item The relative angles $\Delta$ between the current state and the new
	measurement axis form a uniform grid of $m = 10$ angles from $0$ to
	$\pi/2$ radians.
\item Every combination of $\phi$ and $\Delta$ is run once, for $N m = 300$
	trials in a single suite. The 30 values of $\phi$ are visited in a
	shuffled order (the random seed is recorded in the log), so that any slow
	drift over the run is not correlated with $\phi$. For each value of
	$\phi$, the 10 angles are run in sequence. The suite opens with one
	force-off window, so that the first trial also starts from a recorded
	state.
\item After $\phi$ is changed, the controller waits $250$\,ms for the $\phi$
	filter to settle; after the axis is set, it waits $50$\,ms for the $c$ and
	$s$ filters to settle.
\item The force-on and force-off windows each contain 20 readings of
	$(c, s, \theta, \phi, x, y)$ taken every $dt = 50$\,ms (54\,ms in practice,
	including the time to transmit each reading), so a trial lasts about
	$2.2$\,s.
\end{itemize}
Both the Arduino controller code for recording data and the Python
data-analysis scripts are publicly available through GitHub at
\url{https://github.com/DelPhi-kachow/ModelQubit.git}.

We ran this suite once, recording the data through the controller's serial
port; the 300 trials took 11.5 minutes. A Python script
(\texttt{mqAnalysis.py}) splits the recording into trials at each rising edge
of the force. For each trial it computes the realized relative angle $\Delta$
between the new axis and the state just before the force switched on, and the
normalized hidden variable:
\begin{align*}
w = \frac{\phi - \phi_\text{min}}{\phi_\text{max} - \phi_\text{min}} \in [0, 1],
\end{align*}
where $[\phi_\text{min}, \phi_\text{max}]$ is the sampling interval. The
outcome is $\uparrow$ ($\downarrow$) if the final state lies within
$20^\circ$ of $(c, s)$ ($-(c, s)$). A final state farther than that from both
poles means the collapse did not finish in time; such a trial is labeled
ambiguous and counted as half $\uparrow$ and half $\downarrow$. The ideal model
predicts $\uparrow$ exactly when $w < P(\uparrow) = \cos^2(\Delta/2)$.

\begin{figure*}[t]
\centering
\includegraphics[width = 0.8\textwidth]{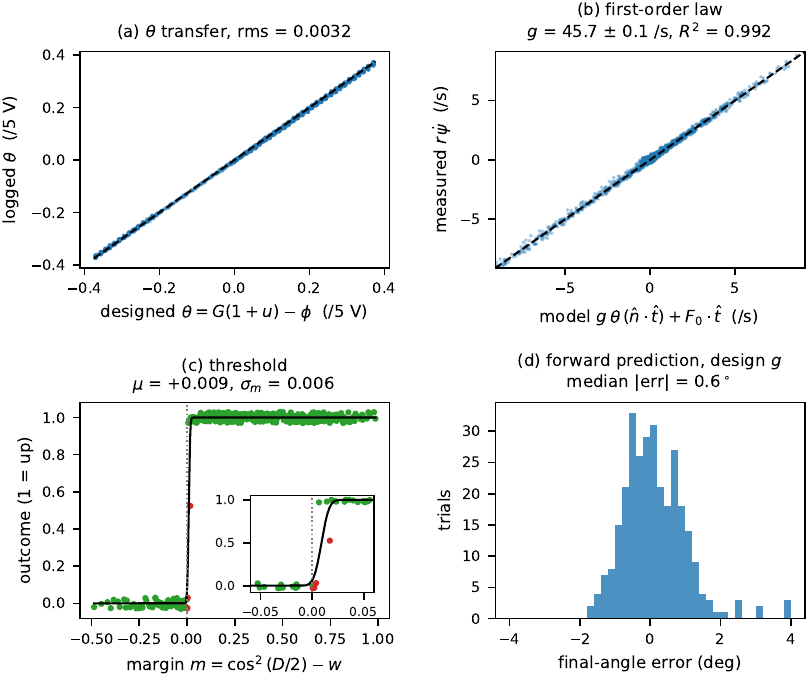}
\caption{Qubit circuit performance against ODE model: (a) Logged $\theta$
	against the designed $\theta = G(1 + u) - \phi$, with $G = 10/47$, for every
	force-on reading; the dashed line is equality. (b) Measured tangential
	velocity $r\dot\psi$ against the fitted first-order law. (c) Outcome
	($1 = \uparrow$) against the margin $m = \cos^2(\Delta/2) - w$ from the
	ideal threshold, with the fitted threshold curve; green points agree with
	the ideal model and red points do not. The inset enlarges the region
	near $m = 0$ (points are jittered vertically for visibility). (d) Final
	angle predicted by integrating the first-order law with the design gain,
	minus the measured final angle.
	}
\label{fig:odeChecks}
\end{figure*}

\subsection{Does the qubit circuit implement the model?}
\label{ssec:odeChecks}

Before looking at probabilities, we check the circuit against the model
itself. Each check compares a quantity measured from the recording with the
value designed into the circuit diagrams of Subsection \ref{ssec:fullCirc}.
To judge whether a deviation is acceptable, we propagated the tolerances of
the components ($1\%$ resistors, $10\%$ capacitors, and the AD633
datasheet specifications) through the circuit diagrams by Monte Carlo
(Appendix \ref{app:tolerance}), giving the band within which $95\%$ of
circuits built from these parts would fall. Table \ref{tab:odeChecks}
summarizes the results, and Fig. \ref{fig:odeChecks} shows the data behind
them. All voltages here are in the logged units, fractions of $5$\,V.

\begin{table}[h]
\caption{Circuit checks: Design values, measured values, and the central
	$95\%$ band of the deviation expected from component tolerances with
	typical AD633 offsets (Appendix \ref{app:tolerance}). Errors on measured
	values are bootstrap standard errors over trials.}
\label{tab:odeChecks}
\begin{ruledtabular}
\begin{tabular}{lcccc}
Quantity & Design & Measured & Dev. & Tolerance band \\
\hline
$a_0$ & $0.2128$ & $0.2116(3)$ & $-0.5\%$ & $\pm 3.6\%$ \\
$a_u$ & $0.2128$ & $0.2098(2)$ & $-1.4\%$ & $\pm 1.8\%$ \\
$a_\phi$ & $-1$ & $-1.0019(11)$ & $+0.2\%$ & $\pm 2.1\%$ \\
$R_\text{eq}$ & $0.8584$ & $0.8655$ & $+0.8\%$ & $\pm 1.0\%$ \\
$g$ (s$^{-1}$) & $50$ & $45.7(1)$ & $-8.6\%$ & $-9.2\%$, $+11.0\%$ \\
$\mu$ & $0$ & $+0.009(3)$ & & $-0.011$, $+0.013$ \\
\end{tabular}
\end{ruledtabular}
\end{table}

\paragraph{$\theta$ circuit.}
During the force-on windows, we fit the logged $\theta$ to
$\theta = a_0 + a_u u + a_\phi \phi$, where $u = xc + ys$. The design
(Subsection \ref{sssec:scaling}) is $a_0 = a_u = 10/47$ and $a_\phi = -1$.
The fit has $R^2 = 0.9998$ with an rms residual of $0.003$
[Fig. \ref{fig:odeChecks}(a)], and all three coefficients fall within their
tolerance bands.

\paragraph{Radius.}
The settled radius is $R_\text{eq} = 0.866$, $0.8\%$ above the designed
$\sqrt{14/19}$, again within tolerance.

\paragraph{Dynamics.}
The drag in the double integrators is strong enough that the velocity stages
follow the forcing almost instantly, so the motion around the circle obeys a
first-order law. Writing the state as $r(\cos\psi, \sin\psi)$ with unit
tangent $\hat{t} = (-\sin\psi, \cos\psi)$ and measurement axis $\hat{n} =
(c, s)$, gives:
\begin{align*}
r \dot{\psi} = g\, \theta\, (\hat{n} \cdot \hat{t}) + F_0 \cdot \hat{t} .
\end{align*}
The measurement gain $g$ is set by the AD633 scale factor and the time
constant of the position integrators ($1\,\mathrm{M}\Omega \times 10$\,nF
$= 10$\,ms), giving a design value of $50$\,s$^{-1}$; $F_0$ is a small
constant force that captures drift. Fitting this law to consecutive pairs of
readings gives $g = 45.7 \pm 0.1$\,s$^{-1}$ with $R^2 = 0.992$ during the
force-on windows [Fig. \ref{fig:odeChecks}(b)]. The $8.6\%$ shortfall is
inside the tolerance band, which is dominated by the $10\%$ tolerance of the
integrator capacitors. The drift is $F_0 = (-0.048, 0.088)$\,s$^{-1}$, about
$0.1$\,s$^{-1}$ in magnitude, compared with forcing terms of several
s$^{-1}$ while the force is on. As an end-to-end test of the dynamics, we
integrate this law with the \emph{design} gain and no drift over each
force-on window, starting from the recorded state and driven by the recorded
$\theta$. The predicted final angle differs from the measured one by a median
of $0.6^\circ$ [Fig. \ref{fig:odeChecks}(d)].

\paragraph{Sign rule.}
The model says the state collapses to $\mathrm{sign}(\theta)\,(c, s)$, with
$\theta$ evaluated when the force switches on. This holds in 299 of the 300
trials. The single exception (trial 183) began with $\theta = -0.002$,
essentially zero, and its collapse did not finish: it is the one ambiguous
trial in the run.

\paragraph{Outcomes versus ideal model.}
Finally, we compare each outcome with the ideal model's prediction for the same
$\Delta$ and $\phi$. They agree in 296 of the 300 trials. To see why the other
four disagree, define each trial's margin from the ideal threshold,
$m = \cos^2(\Delta/2) - w$, so that the ideal model says $\uparrow$ exactly when
$m > 0$. All four disagreements have $|m| \le 0.017$, right at the threshold
[Fig. \ref{fig:odeChecks}(c)]. A maximum-likelihood fit of
$P(\uparrow) = \Phi\big((m - \mu)/\sigma_m\big)$, where $\Phi$ is the standard
normal cumulative distribution function, gives a threshold bias $\mu = +0.009$
(68\% interval $+0.006$ to $+0.012$) and a threshold jitter $\sigma_m = 0.006$,
both in units of probability. The bias is within the band expected from the
AD633 offsets. In other words, the circuit reproduces the ideal model's
decisions to within one percent of the threshold.

\begin{figure}[htbp]
\centering
\includegraphics[width=\columnwidth]{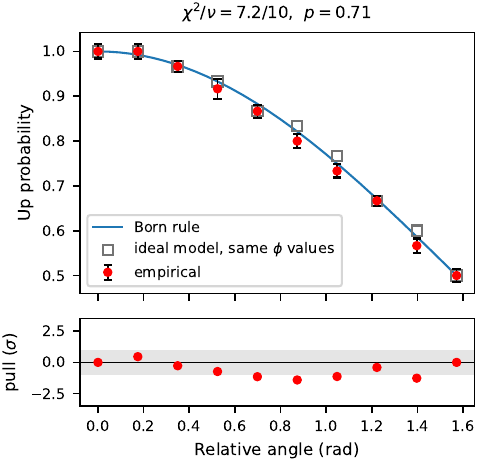}
\caption{Empirical probability of collapsing to $\uparrow$ against the relative
	angle between the state and the measurement axis: (top) Measured
	probabilities (red, with error bars from the error model of
	Appendix \ref{app:errorModel}) compared with the Born Rule (blue line) and
	with what a perfect circuit gives for exactly the same values of $\phi$
	(open squares). (bottom) Pull, $(\hat{p} - P_\text{Born})/\sigma$, for each
	relative angle; the shaded band is $\pm 1\sigma$.
	}
\label{fig:empiricalProb}
\end{figure}

\subsection{Born Rule statistics}
\label{ssec:born}

For each relative angle, the empirical probability $\hat{p}$ of collapsing to
$\uparrow$ is the fraction of its 30 trials that ended at $\uparrow$ (with the
ambiguous trial counted as one half). It is compared with the Born Rule
probability $\cos^2(\Delta/2)$ averaged over the realized angles of those
trials. The results are plotted in Fig. \ref{fig:empiricalProb}. More details,
including an animation of all 300 trials collapsing and diagnostic animations
of individual trials, are available on the aforementioned GitHub repository.

The error bars come from an error model built from the data (Appendix
\ref{app:errorModel}). It has three parts. The first is the sampling of the
hidden variable: with 30 values of $\phi$, each empirical probability is a
30-point estimate of an integral over $\phi$, precise to about $1/(2N) \approx
0.017$. The second is the circuit's own threshold noise, taken from the fit
of $\mu$ and $\sigma_m$ above. The third accounts for ambiguous trials. The
resulting error bars range from $0.011$ to $0.022$. Monte Carlo simulation of
the model reproduces the spread of $\hat{p}$ to within a few percent in every
bin.

Against these error bars, the empirical probabilities agree with the Born Rule
with $\chi^2 = 7.2$ for 10 degrees of freedom ($p = 0.71$), and the rms
difference between $\hat{p}$ and the Born Rule is $0.013$. For comparison, a
perfect circuit run with exactly the same 30 values of $\phi$ would differ from
the Born Rule by $0.010$ rms, so most of the remaining difference comes from
the finite sampling of $\phi$ rather than from the circuit. We conclude that
the analog qubit recreates Born Rule statistics to within about $0.02$ in
probability.

Since the circuit is deterministic, a departure from the Born Rule enters
only through a few identifiable routes, each of which the checks above
constrain. A bias $\mu$ in the decision threshold shifts every probability by
about $\mu$; the fitted $\mu = +0.009$ is below the size of the error bars. A
distribution of $\phi$ that is not uniform over its sampling interval, or a
$\theta$ circuit that responds to $\phi$ nonlinearly, would distort
$P(\uparrow)$ into a different function of the angle; the $\phi$ values here
are set on a uniform grid and read back, and the fitted $\theta$ depends
linearly on $\phi$ with a coefficient within $0.2\%$ of design. Finally, a
$\phi$ that changed during a collapse could flip the sign of $\theta$ partway
through; the sign rule holds in 299 of 300 trials. None of these routes is
detectable in the present data. Conversely, introducing them on purpose, for
example by drawing $\phi$ from a nonuniform distribution, would produce
controlled departures from the Born Rule, which we leave to future work.

\subsection{Remaining limitations}
\label{ssec:limitations}

An analog circuit implementing a measured qubit brings with it its own
limitations. Let's address four of these:
\begin{enumerate}
     \setlength{\topsep}{-5pt}
      \setlength{\itemsep}{-5pt}
      \setlength{\parsep}{-5pt}
      \setlength{\labelwidth}{5pt}
\item \textit{Resolution of the hidden variable.} With 30 values of $\phi$,
	the empirical probabilities cannot resolve the Born Rule more finely than
	about $0.017$. The number of values is limited by run time: the circuit's
	time scale was chosen so that a trial takes a couple of seconds and can be
	followed on an oscilloscope in real time, both for troubleshooting and as a
	visual demonstration of the computation. The resolution of the PWM drive
	for $\phi$ (about 186 levels across the sampling interval) would allow a
	somewhat denser grid.

	In practice, one swaps out resistors and capacitors in the double
	integrators to run the experiment much faster. The LF411 op-amps
	\cite{lf411} used in our analog computer have a slew rate high enough to
	handle signals into the audio range. The upshot is that we can, easily, run
	the model 100 times faster by swapping out the $\SI{1}{\mega \ohm}$
	resistors in the double integrators with $\SI{10}{\kilo \ohm}$ resistors.

\item \textit{Drift.} With the force off, the state is not perfectly at rest;
	it drifts at about $0.1$\,s$^{-1}$ (the fitted $F_0$), which can be seen in
	the force-off windows of Fig. \ref{fig:trialCycle}. Since each trial's
	relative angle is measured from the state just before the force switches
	on, the drift does not bias the comparison with the Born Rule. And, while
	the force is on, it is small compared with the measurement forcing. The
	drift likely arises from small asymmetries between the $x$ and $y$
	channels, such as the $10\%$ tolerance of the integrator capacitors and
	the multiplier offsets, which are amplified by the large radial force
	constant $k$. Weakening the radial force would reduce the drift, at the
	expense of letting the state stretch off the circle.

\item \textit{Incomplete collapse.} When $\theta$ is very close to zero, the
	measurement force is weak and the collapse may not finish within the
	force-on window. This happened once in 300 trials (trial 183), which was
	counted as half $\uparrow$ and half $\downarrow$ and contributes to the
	error bar of its bin.

\item \textit{Offsets.} The threshold bias $\mu = +0.009$ and the $8.6\%$
	shortfall in the measurement gain are both within the tolerances of the
	parts. The gain affects only how fast the state collapses, not where it
	goes. The bias shifts every threshold by about one percent and is the
	largest systematic effect on the outcomes; trimming the AD633 offsets
	would reduce it.
\end{enumerate}

%% file: discussion/discussion.tex
\section{Discussion}
\label{sec:discussion}

We demonstrated how to design a deterministic dynamical system that models
measurement in a single qubit. Our electronic analog computer implementation
confirmed that this model recreates the statistics of the Born rule to within
about $0.02$ in probability and that the circuit reproduces the ideal model's
individual outcomes to within one percent of the decision threshold. In this
way, the implementation illustrated the potential of deterministic collapse
theories, which explain measurement with a deterministic dynamical system that
modifies the Schr\"odinger equation with nonlinear terms. Beyond the
theoretical implications, the analog implementation itself demonstrates the
ability of deterministic dynamical systems to replicate key aspects of the
behavior of quantum systems. More exploration is needed to probe how practical
measured analog qubits will be for our eventual target of 100s or 1000s
of them coupled on VLSI substrates.

Before touching on the physical implications, we should discuss the limitations
of our model. Although the model has the structure required of a dynamical
model of measurement (Subsection \ref{sssec:modelSummary}), the particular
equations are ad hoc, and we certainly do not offer them verbatim as a final
theory of measurement. In particular, we recognize that the model's $\theta$
term, defined in Eq. (\ref{eqn:thetaDef}) to decide the measurement outcome
via its sign, only makes sense for a system with two distinct measurement
outcomes (a qubit). Furthermore, the construction of Eqs. (\ref{eqn:c1}) -
(\ref{eqn:c2}) is constrained to be easily implemented on an analog computer,
using only sums, differences, and products in the equations of motion. This
constraint is not motivated physically, but chosen for ease of construction.
There are no doubt similar models that also recreate Born rule statistics but
use terms that would not be easily implemented as an analog circuit. However,
these alternative constructions would still fall under the umbrella of
``deterministic collapse theories''.

The particular equations of motion aside, the truly salient feature of our
model is that it is a deterministic nonlinear system that reduces to
Schr\"odinger evolution when a measurement is not happening, but creates
Born-rule statistics when a measurement occurs. Measurement is produced
by nonlinear time evolution, which is interpreted as a superposition-violating
Hamiltonian. This interpretation is tautologically correct if the Hamiltonian
is defined as the generator of time evolution rather than a physical notion of
energy. Intuitively, this makes sense, as a measurement must be able to
distinguish between an arbitrary superposition of basis states and a pure
eigenstate of the measurement. All deterministic dynamical collapse theories
posit a continuous interpolation between unitary (Schr\"odinger) and
nonunitary (measurement) dynamics. This aspect is shared with the continuous
spontaneous localization theory \cite{CSL} and the Di\'osi-Penrose theory
\cite{DP}, though deterministic collapse differs from these by not positing a
source of intrinsically random noise.

The model's state is classical. An agent with direct access to the circuit's
voltages could read out $(x, y)$, or copy it, as anyone probing the circuit
does. The model is meant to describe the qubit as seen by an observer whose
only access to it is through measurement: switching on the measurement force
and recording the outcome. With that restriction, the model respects the
single-qubit counterpart of no-cloning, that nonorthogonal states cannot be
reliably distinguished from one measurement. The outcome depends on the state
only through $P(\uparrow)$, compared against the unresolved $\phi$, so one
measurement reveals no more about the state than a projective measurement in
standard quantum mechanics. This is not automatic. A measurement force that
ignored $\phi$, such as $\theta = xc + ys$, would distinguish nonorthogonal
states deterministically. Requiring the outcome to depend on the state only
through its Born probabilities is therefore a real constraint on the allowed
measurement dynamics. In our model, it stands in for the protection that
linearity provides in standard quantum mechanics.

Our model and experiment say nothing about where the hidden variable $\phi$
comes from; they require only that it be uniformly distributed from the
observer's point of view and that it hold still while a measurement completes.
By way of looking to future efforts, let's discuss various options to these
constraints.

One natural way to meet both requirements is for $\phi$ to be a fast,
deterministic oscillation whose phase the observer cannot resolve. A triangle
wave spanning $[0, 2]$, for instance, spends equal time at every value, so an
observer who triggers a measurement at an imprecisely known time sees a
uniformly distributed $\phi$, provided the period is long compared with the
collapse time but short compared with the observer's timing uncertainty. It is
then suggestive to imagine that this oscillation corresponds to the global
phase of the qubit, which would oscillate very fast, generating apparent
randomness. For instance, oscillation due to an energy difference on the order
of $1$ \unit{eV} would be on the order of $200$ \unit{THz}. If the measurement
apparatus were treated as part of the system under study, we would instead be
tempted to interpret $\phi$ as a relative phase between the qubit and the
measurement apparatus. One motivation to expect this is that nonlinear terms in
a Hamiltonian with respect to $\sPsi = A\sZ + B\sO$, proportional to $A^n$ or
$B^n$, would oscillate faster with respect to the phase of $\sPsi$ than linear
terms (a linear combination of $A$ and $B$). Phase as the hidden variable of
dynamical collapse becomes testable when oscillations occur at observable time
scales. 

Dynamical collapse theories are hidden-variable theories, so Bell's theorem
(Sec.~\ref{sec:intro}) applies to them. Two points keep it from ruling out
our model. First, Bell's theorem constrains only correlations between two or
more systems. A single qubit, which is all our circuit models, admits a
deterministic hidden-variable description; Bell himself constructed one for a
spin-$1/2$ particle \cite{Bell_1966}. The theorem applies only once the model is
extended to entangled pairs. Second, in such an extension the collapse
dynamics of each qubit would depend on the joint state, making dynamical
collapse a \emph{nonlocal} hidden-variable theory, which is not bound by
Bell's inequality. (The other escape route, abandoning Bell's assumption of
statistical independence \cite{supermeasured}, is not needed here.) Standard
quantum mechanics is nonlocal in the same way; observers in different
reference frames will have the same measurement outcomes, but they will
disagree about the causal ordering of states and collapses.

This curt answer to entanglement is not enough to prove consistency with
causality; more investigation is still needed. In particular, we are developing
an explicit dynamical collapse model of a Bell test to explore this. Dynamical
collapse adds new meaning to the state of a system, because it is now also
defined (and continuously evolving) during a measurement. It is desirable to
have an account of state trajectories in different reference frames and a
transformation rule between them. More work is needed to definitively show that
dynamical collapse is consistent with causality. However, it is safe to say
that Bell's Theorem does not rule out dynamical collapse theories from the
outset. 

This model is similar in spirit to that of Weinberg \cite{Weinberg1989}, who
used observables that are nonlinear functions of the state. Gisin and
Polchinski showed that Weinberg's nonlinear dynamics permits superluminal
signaling \cite{Gisin1989, Gisin1990, Polchinski1991}. Linearity is the
standard guarantee against this. By the no-cloning argument of
Sec.~\ref{ssec:measProblem}, linear dynamics cannot copy an unknown state, and
so cannot be used to signal through entanglement \cite{Dieks1982}. Linearity
is sufficient for no-signaling, however, not necessary. Gisin's argument
assumes that the nonlinear evolution depends only on the quantum state. A
distant party's choice of measurement then prepares different ensembles of
pure states with the same density matrix, and the nonlinear evolution tells
them apart. Our class of models does not satisfy that assumption: its dynamics
also depend on the hidden variable $\phi$, which standard quantum mechanics
does not represent. (Above, we speculated that $\phi$ may be a global phase.)

Pilot-wave theory shows that such an escape is possible in principle. Its
guidance equation is nonlinear and nonlocal, yet it permits no signaling as
long as particle positions are Born-distributed, a condition known as quantum
equilibrium. Valentini showed that this holds in any deterministic
hidden-variable theory: signaling is impossible in equilibrium and generically
possible outside it \cite{Valentini2002}. Our model is such a theory, and the
requirement that $\phi$ be uniformly distributed from the observer's point of
view plays the role of equilibrium. We therefore expect no-signaling in a
multi-qubit extension of the model, if it can be achieved, to rest on that
requirement. Because cloning would itself permit signaling \cite{Dieks1982},
establishing no-signaling for entangled qubits would also settle whether the
model admits cloning. This is a central question for the dynamical-collapse
Bell test described above.

Bieli\'nska,
Eckstein, and Horodecki recently re-examined Gisin's no-go theorem in light of
later developments, clarifying which of its assumptions a nonlinear model must
evade to avoid superluminal signaling. They showed that such nonlinear
dynamics is generically chaotic \cite{Bielinska2025}. Dependence on a variable
hidden from standard quantum mechanics is one candidate route through these
loopholes, though more work is needed to be certain.

The literature on weak measurements provides further insight into how deterministic collapse models ought to be refined. Patel and Kumar show that imposing continuous satisfaction of the Born rule is sufficient for satisfying relativity, and they give stochastic models of time evolution that satisfy this \cite{weak}. On the experimental side, Murch et al. use a state tomography technique to ``watch" the state of a superconducting qubit embedded in a microwave cavity as it collapses under weak measurement \cite{singleTrajectories}. These observations show the state undergoing what appears to be a random walk towards an eigenstate of the measurement. Qualitatively, this means that theoretical models of measurement should not give ballistic trajectories; they should model the noise. This poses a challenge to deterministic collapse models, which will likely require modeling nonlinear coupling of the state to the environment in addition to between the state and the measurement device.  

In addition to harmonizing dynamical collapse with causality and modeling the noisiness of weak measurement experiments, there are several other paths to follow on the theory side:
\begin{enumerate}
     \setlength{\topsep}{-5pt}
      \setlength{\itemsep}{-5pt}
      \setlength{\parsep}{-5pt}
      \setlength{\labelwidth}{5pt}
\item A dynamical collapse model that accounts for more than two discrete 	states.
\item A dynamical collapse model for a continuous basis, such as with position 	or momentum.
\item Investigation into other hidden variables as sources of sensitive dependence on initial conditions. We posited global phase (or relative phase between object and measurement device). Besides variables that are completely hidden to quantum theory, there are other variables that are
only ``slightly hidden'', particularly states in the measurement device and states in the environment, which are still traced out when constructing the system's reduced density matrix. 
\item A dynamical collapse model in QFT, which would likely come out as an extension of causality, as in QFT.
\item Devising concrete experimental tests for intermediate regimes between unitary evolution and measurement.
\item Devising concrete experimental tests for particular hidden variable dependence.
\item Measurement of two entangled qubits. This requires designing measurement forces that act on a joint state, and it is also the setting for an explicit dynamical-collapse model of a Bell test.
\item Chaos. The present model collapses onto fixed points. Whether the framework can produce chaotic dynamics, and what that would mean for the corresponding quantum system, is open; nonlinear quantum dynamics is known to be generically chaotic \cite{Bielinska2025}.
\item Robustness to noise. Adding Langevin noise to the equations of motion, or to the circuit, and measuring how the Born-rule frequencies degrade would connect deterministic collapse to the stochastic collapse theories and to the noisy trajectories seen in weak measurement.
\item A geometric formulation. Our model restricts the qubit to real amplitudes, a great circle of the Bloch sphere. Geometric quantum mechanics describes states as points and distributions on complex projective space \cite{Anza2021}, and would be a natural language for extending deterministic collapse to the full state space.
\item Speed limits: Bounding how fast collapse can proceed. Our model is a dissipative deterministic system whose collapse rate is set by the measurement gain and drag, making it a natural test case for recent speed limits on deterministic dynamical systems, which bound the rate of change of observables using a classical analog of the density matrix and Fisher information on the tangent-space dynamics \cite{Das2022, Das2023}.
\end{enumerate}
In view of these, our construction of an analog qubit appears to open up a
number of empirical and theoretical challenges.

\section{Conclusion}
\label{sec:Conclusion}

Fundamental physics aside, these results may have applications in the realm of computer engineering. We demonstrated that a single classical dynamical system can reproduce the behavior of a measured qubit: The same equations describe both nonmeasurement evolution and continuous collapse when the measurement coupling is turned on, and the outcomes follow Born-rule statistics. This is less than full quantum behavior. The model's state is classical information, and whether the dynamics can be extended to entangled qubits without permitting cloning or signaling remains open (Sec.~\ref{sec:discussion}). As mentioned in the introduction, such systems could be useful for efficiently simulating quantum algorithms on classical
systems. However, we have not demonstrated here that such analog computers can easily construct high-dimensional Hilbert spaces, which is the same problem encountered by other attempts to create (nonmeasurement) analogs to quantum computing \cite{Kish2003, Mourya_CMOS}. 

To understand the scalability problem more deeply, we need to contrast the amount of hardware needed for a quantum computer against our current approach to constructing an analog simulation. In a real quantum computer, one constructs a number $N$ of qubits, then runs those qubits through a quantum circuit, possibly taking measurements throughout the process. The state space of these $N$ qubits is $2^N$-dimensional, with each dimension being a complex number. Our analog emulation approach uses a separate voltage to model each real-valued parameter. It takes two real variables, $x$ and $y$, to model each complex variable of the form $z = x + iy$. Thus, our current approach to analog simulation requires $2^{N+1}$ subcircuits to model the same $N$ qubits. This results in an exponential growth in the number of component parts, making simulation of large quantum computers infeasible.

\begin{figure*}
\includegraphics[width=\textwidth]{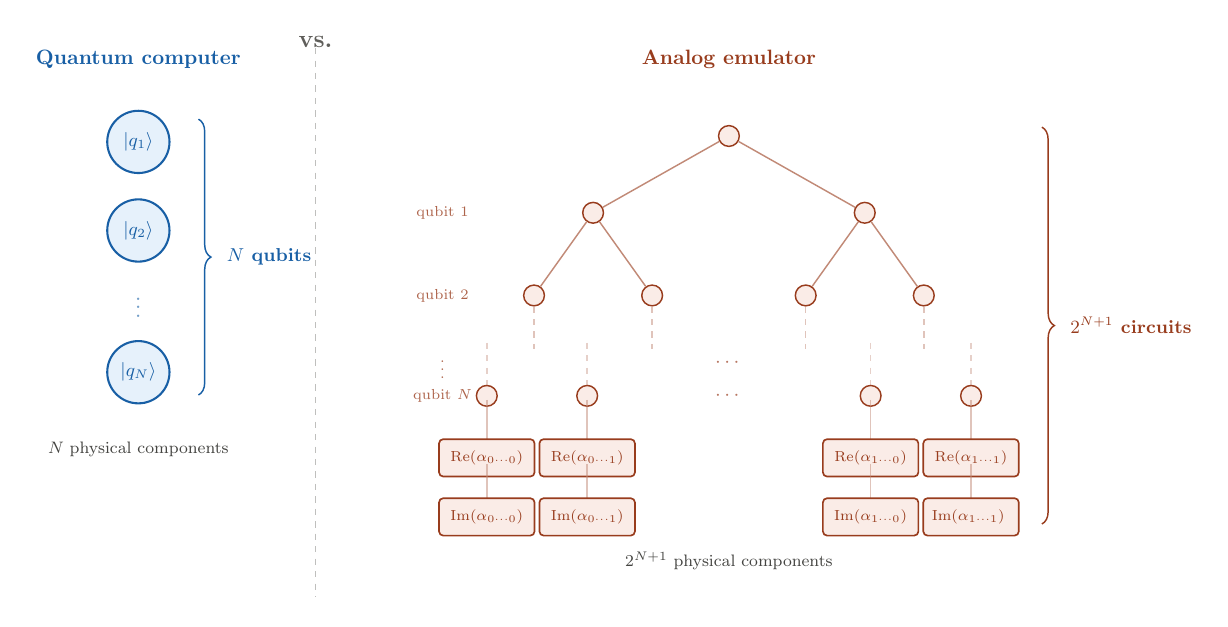}
\caption{Exponential blow-up of components needed to emulate $N$
qubits using our current approach.}
\label{fig:scalability}
\end{figure*}

This blow-up is depicted in Fig. \ref{fig:scalability}. Suppose that, by some
technological miracle, we could construct a circuit that evolved a mole of real
variables ($\approx 6 \times 10^{23}$). Such a device could only simulate 77
qubits. A more reasonable achievement would be modeling a billion parameters,
which allows emulation of 19 qubits. This is not enough to achieve the
much-renowned ``quantum supremacy" on an analog computer. Then again the path
to achieving this would likely be instructive for resource scaling.

Solving the scalability problem would require designing a classical analog to
the tensor product, somehow allowing the implicit creation of new degrees of
freedom through interactions between subsystems (like entanglement), rather
than having to explicitly construct all degrees of freedom beforehand.

Intuition that such systems may exist comes from entropy creation and
bifurcations in dynamical systems theory. For instance, the number bits needed
to accurately predict the chaotic Baker's Map grows with the number of time
steps \cite[p. 464]{strog}. Each new bit can be thought of as a new (binary)
variable, though the only physical form of storage is a single real number
undergoing nonlinear time evolution.

Additionally, it is notable that probability distributions of coupled variables
already have a tensor product structure, which may prove a fruitful direction
of inquiry. At the macroscopic level, something like a fluid can be treated
this way, as a time-evolving ensemble. However, approaches like these are
likely to face a roadblock in that the continuum hypothesis eventually breaks
down at some level of granularity, and a material substance has only a list of
continuous variables that grow linearly with the number of particles, unlike
quantum mechanics. Such limitations are discussed in \cite{montina08,
montina11}.

A scalable classical analog would also open a new class of benchmarks. In many-qubit circuits, interleaving measurements with unitary gates drives a transition from extensive (volume-law) to bounded (area-law) entanglement as the measurement rate increases \cite{Skinner2019}, and such measurement-induced phase transitions have now been observed on quantum hardware \cite{Noel2022}. Because measurement is precisely the ingredient our model adds to earlier analog simulators, these transitions would be a natural test for an analog emulator of many measured qubits.

In summary, this analog simulation of measuring a single qubit motivates a
broader research program with two distinct high-payoff directions. For the
foundations of quantum mechanics, this work revives interest in nonlinear
dynamical systems as a framework for attacking the measurement problem, though
causality problems remain a significant obstacle. On the engineering side,
classical analog emulation of quantum systems hints at the possibility of
efficiently simulating quantum algorithms without quantum hardware, though the
scalability problem is as yet unsolved and would require a genuinely novel
classical analog of the tensor product to overcome. Either of these directions,
if successful, would have profound implications; this program gains resilience
from the fact that failure in one need not preclude success in the other.

%% file: appendix/appendix.tex
\appendix

\section{Empirical probability errors}
\label{app:errorModel}

This appendix derives the error bars of Fig. \ref{fig:empiricalProb}. They
represent the spread of the empirical probabilities expected if the same
circuit were run again with the same protocol.

Trial $i$ has realized relative angle $\Delta_i$, normalized hidden variable
$w_i \in [0, 1]$, and Born probability $P_i = \cos^2(\Delta_i/2)$. The ideal
model gives $\uparrow$ exactly when the margin $m_i = P_i - w_i$ is positive.
For the bin of relative angle $j$, containing $n_j = 30$ trials of which
$n_{\text{amb},j}$ are ambiguous, the empirical probability is:
\begin{align*}
\hat{p}_j = \frac{n_{\uparrow, j} + \tfrac{1}{2} n_{\text{amb}, j}}{n_j} ,
\end{align*}
and it is compared with $P_j$, the mean of $P_i$ over the bin. Using the mean
over the realized angles means that any difference between the commanded and
realized angles is accounted for in $P_j$ rather than added to the error bar.

\paragraph{Device threshold noise.}
A real circuit does not switch exactly at $m = 0$. We model each outcome as
$\uparrow$ with probability $\Phi\big((m_i - \mu)/\sigma_m\big)$, where $\Phi$
is the standard normal cumulative distribution function, $\mu$ is the
circuit's threshold bias, and $\sigma_m$ its threshold jitter, both in units of
probability. The parameters are fit by maximum likelihood over all 300 trials,
giving $\mu = +0.009$ and $\sigma_m = 0.006$ (Subsection \ref{ssec:odeChecks}).

\paragraph{Hidden-variable sampling.}
The hidden variable takes only $N = 30$ values, so $\hat{p}_j$ is a 30-point
quadrature of a step function of $w$ rather than an exact integral. To model
the resulting uncertainty, we treat each value of $\phi$ as a single uniform
draw $w_c + U$ from its grid cell of width $h = 1/N$, where $w_c$ is the cell
center and $U$ is uniform on $[-h/2, h/2]$. (A repeated suite would reuse the
same values of $\phi$, so it would share $U$ and add no new samples of the
hidden variable.) Given $U$, trial $i$ in cell $c$ is $\uparrow$ with
probability:
\begin{align*}
p_i(U) = \Phi\!\left( \frac{P_i - w_c - U - \mu}{\sigma_m} \right) .
\end{align*}

\paragraph{Combining the two.}
By the law of total variance, the variance of the number of $\uparrow$
outcomes in each cell splits into a part from the hidden variable and a part
from the circuit. Summing over the cells $c$ in bin $j$,
\begin{align*}
\sigma_{\text{HV}, j}^2 &= \frac{1}{n_j^2} \sum_c \mathrm{Var}_U
	\Big[ \sum_{i \in c} p_i(U) \Big] , \\
\sigma_{\text{dev}, j}^2 &= \frac{1}{n_j^2} \sum_c \mathrm{E}_U
	\Big[ \sum_{i \in c} p_i(U) \big(1 - p_i(U)\big) \Big] ,
\end{align*}
where the sums over $i$ run over the trials of bin $j$ in cell $c$. Each
ambiguous trial, counted as one half, adds $\sigma_{\text{amb}, j}^2 = 0.25\,
n_{\text{amb}, j} / n_j^2$. The error bar is:
\begin{align*}
\sigma_j^2 = \sigma_{\text{HV}, j}^2 + \sigma_{\text{dev}, j}^2
	+ \sigma_{\text{amb}, j}^2 ,
\end{align*}
with a floor of $0.002$. The floor matters only where the grid of $\phi$ is
symmetric about the threshold, as at $\Delta = 90^\circ$, where the other terms
nearly vanish. For this run $\sigma_{\text{HV}}$ and $\sigma_{\text{dev}}$ are
each about $0.01$, the one ambiguous trial adds $0.017$ to its bin, and the
combined error bars range from $0.011$ to $0.022$. A
Monte Carlo simulation of the model reproduces the spread of $\hat{p}_j$ to
within a few percent in every bin.

\paragraph{Bias of the grid.}
The same framework gives the expected bias of $\hat{p}_j$ for a perfect
circuit, $\mathrm{E}_U[\hat{p}_j] - P_j$ with $\sigma_m \to 0$ and $\mu = 0$.
For a grid that includes the endpoints of the sampling interval, this bias
pulls the estimate toward $1/2$ by up to about $1/(2N)$. For the midpoint grid
used here it is negligible: the values computed for this run are at most
$0.003$ in magnitude.

\section{Component-tolerance Monte Carlo}
\label{app:tolerance}

The tolerance bands in Table \ref{tab:odeChecks} come from a Monte Carlo
simulation (\texttt{toleranceMC.py} in the GitHub repository) that propagates
component tolerances through the circuit diagrams of Subsection
\ref{ssec:fullCirc} to the same quantities the analysis measures, in the same
logged units.

Each of $40{,}000$ simulated circuits draws every resistor uniformly within
$\pm 1\%$ of its nominal value, every capacitor within $\pm 10\%$, and each
AD633 scale factor within $\pm 0.25\%$. Each AD633 also receives output and
input offsets, drawn uniformly within the datasheet limits \cite{ad633}. Two
scenarios are run: typical offsets ($\pm 5$\,mV at the output and inputs)
and maximum offsets ($\pm 50$\,mV at the output, $\pm 30$\,mV at the inputs).
All $+5$\,V references, including the Arduino's ADC reference, are treated as
a single rail at its nominal value.

For each simulated circuit, the simulation computes the following, using the
same conversions from pin voltages to logged values as the firmware:
\begin{itemize}
     \setlength{\topsep}{-5pt}
      \setlength{\itemsep}{-5pt}
      \setlength{\parsep}{-5pt}
      \setlength{\labelwidth}{5pt}
\item The $\theta$ coefficients $a_0$, $a_u$, and $a_\phi$, by evaluating the
	$\theta$ circuit (including the drive and readout stages) over a grid of
	states, axes, and values of $\phi$, and fitting the same linear model used
	on the data.
\item The equilibrium radius, by solving for the voltage at which the radial
	force vanishes in each channel and averaging the two channels.
\item The measurement gain $g$, from the small-signal gain of the chain from
	$\theta$ through the $\theta c$ multiplier and the velocity and position
	stages to the logged $x$.
\item The threshold bias $\mu$, from the value of $\phi$ at which the
	simulated $\theta$ changes sign at each of the 10 relative angles,
	compared with the ideal threshold $\cos^2(\Delta/2)$.
\end{itemize}
The band for each quantity is the central $95\%$ interval of its deviation
from design. Table \ref{tab:odeChecks} lists the typical-offset bands; with
maximum offsets they widen to $-4.8\%$ to $+4.9\%$ for $a_0$, $\pm 2.4\%$ for
the radius, and $-0.024$ to $+0.030$ for $\mu$, while the other bands are
essentially unchanged. The band for $g$ is dominated by the capacitors of the
position integrators. The supply rail is not included: the slope $a_u$ and the
gain $g$ scale in proportion to it, and the radius scales as its inverse
square root, so a rail a few percent from $5$\,V would shift these quantities
by a comparable amount.

%% file: refs.bib
@article{Thom81a,
  title = {The tide gauge, tidal harmonic analyser, and tide predicter},
  author = {W. Thomson},
  journal = {Proc. Inst. Civil Engineers},
  volume = {65},
  optissue = {},
  pages = {3--24},
  optnumpages = {},
  year = {1881},
  optmonth = {},
}

@misc{normalExp,
      title={Thermodynamic Matrix Exponentials and Thermodynamic Parallelism}, 
      author={Samuel Duffield and Maxwell Aifer and Gavin Crooks and Thomas Ahle and Patrick J. Coles},
      year={2024},
      eprint={2311.12759},
      archivePrefix={arXiv},
      primaryClass={cond-mat.stat-mech},
      url={https://arxiv.org/abs/2311.12759}, 
}

@book{strog,
   title =     {Nonlinear Dynamics and Chaos: With Applications to Physics, Biology, Chemistry, and Engineering},
   author =    {Steven H. Strogatz},
   publisher = {Westview Press},
   isbn =      {0813349109; 9780813349107},
   year =      {2014},
   series =    {Studies in Nonlinearity},
   edition =   {2nd},
}

@book{practical,
   title =     {Practical Electronics for Inventors},
   author =    {Scherz, Paul and Monk, Simon},
   publisher = {McGraw-Hill Education TAB},
   isbn =      {9781259587542; 1259587541},
   year =      {2016},
   edition =   {4th Edition}}

@book{mims,
   title =     {Timer, Op Amp \& Optoelectronic Circuits \& Projects},
   author =    {Mims, Forrest M.},
   publisher = {Master Publishing, Inc.},
   isbn =      {978-0945053293},
   year =      {2007},
   edition =   {4th Edition}}

@manual{ad633,
    organization  = {Analog Devices},
    title         = {Low Cost Analog Multiplier},
    number        = {AD633},
    year          =  {2015},
    month         =  {3},    
}

@book{vanValkenburg,
   title =     {Analog Filter Design},
   author =    {Van Valkenburg, M. E.},
   publisher = {Holt, Rinehart and Winston},
   address =   {New York},
   isbn =      {9780030592461},
   year =      {1982}}

@article{sallenKey,
   author =    {Sallen, R. P. and Key, E. L.},
   title =     {A Practical Method of Designing {RC} Active Filters},
   journal =   {IRE Transactions on Circuit Theory},
   volume =    {2},
   number =    {1},
   pages =     {74--85},
   year =      {1955},
   month =     mar,
   doi =       {10.1109/TCT.1955.6500159}}

@manual{lf411,
    organization  = {Texas Instruments},
    title         = {LF411 FET-Input Operational Amplifier},
    number        = {LF411},
    year          =  {2025},
    month         =  {10},    
}

@BOOK{zet,
  title     = "Quantum Mechanics",
  author    = "Zettili, Nouredine",
  publisher = "Wiley-Blackwell",
  edition   =  2,
  month     =  jan,
  year      =  2009,
  address   = "Hoboken, NJ",
}

@misc{analogQubit,
  author = {Muzzin, Jae F.},
  title = {Analog Simulation of the Quantum Bit},
  year = 2024,
  url = {https://medium.com/intuition/analog-simulation-of-the-quantum-bit-de05cbef1a0c},
}

@article{weak,
  title = {Weak measurements, quantum-state collapse, and the Born rule},
  author = {Patel, Apoorva and Kumar, Parveen},
  journal = {Phys. Rev. A},
  volume = {96},
  issue = {2},
  pages = {022108},
  numpages = {11},
  year = {2017},
  month = {Aug},
  publisher = {American Physical Society},
  doi = {10.1103/PhysRevA.96.022108},
  url = {https://link.aps.org/doi/10.1103/PhysRevA.96.022108}
}

@ARTICLE{Mourya_CMOS,

  author={Mourya, Sharan and Cour, Brian R. La and Sahoo, Bibhu Datta},

  journal={IEEE Transactions on Quantum Engineering}, 

  title={Emulation of Quantum Algorithms Using CMOS Analog Circuits}, 

  year={2023},

  volume={4},

  number={},

  pages={1-16},

  doi={10.1109/TQE.2023.3319599}}

@article{supermeasured,
  title = {Supermeasured: Violating Bell-Statistical Independence Without Violating Physical Statistical Independence},
  volume = {52},
  ISSN = {1572-9516},
  url = {http://dx.doi.org/10.1007/s10701-022-00602-9},
  DOI = {10.1007/s10701-022-00602-9},
  number = {4},
  journal = {Foundations of Physics},
  publisher = {Springer Science and Business Media LLC},
  author = {Hance,  Jonte R. and Hossenfelder,  Sabine and Palmer,  Tim N.},
  year = {2022},
  month = jul 
}

@article{Bell_EPR,
  title = {On the Einstein Podolsky Rosen paradox},
  author = {Bell, J. S.},
  journal = {Physics Physique Fizika},
  volume = {1},
  issue = {3},
  pages = {195--200},
  numpages = {6},
  year = {1964},
  month = {Nov},
  publisher = {American Physical Society},
  doi = {10.1103/PhysicsPhysiqueFizika.1.195},
  url = {https://link.aps.org/doi/10.1103/PhysicsPhysiqueFizika.1.195}
}

@article{CHSH,
  title = {Proposed Experiment to Test Local Hidden-Variable Theories},
  author = {Clauser, John F. and Horne, Michael A. and Shimony, Abner and Holt, Richard A.},
  journal = {Phys. Rev. Lett.},
  volume = {23},
  issue = {15},
  pages = {880--884},
  numpages = {0},
  year = {1969},
  month = {Oct},
  publisher = {American Physical Society},
  doi = {10.1103/PhysRevLett.23.880},
  url = {https://link.aps.org/doi/10.1103/PhysRevLett.23.880}
}

@article{superconductingBell,
  title = {Loophole-free Bell inequality violation with superconducting circuits},
  volume = {617},
  ISSN = {1476-4687},
  url = {http://dx.doi.org/10.1038/s41586-023-05885-0},
  DOI = {10.1038/s41586-023-05885-0},
  number = {7960},
  journal = {Nature},
  publisher = {Springer Science and Business Media LLC},
  author = {Storz,  Simon and Sch\"{a}r,  Josua and Kulikov,  Anatoly and Magnard,  Paul and Kurpiers,  Philipp and L\"{u}tolf,  Janis and Walter,  Theo and Copetudo,  Adrian and Reuer,  Kevin and Akin,  Abdulkadir and Besse,  Jean-Claude and Gabureac,  Mihai and Norris,  Graham J. and Rosario,  Andrés and Martin,  Ferran and Martinez,  José and Amaya,  Waldimar and Mitchell,  Morgan W. and Abellan,  Carlos and Bancal,  Jean-Daniel and Sangouard,  Nicolas and Royer,  Baptiste and Blais,  Alexandre and Wallraff,  Andreas},
  year = {2023},
  month = may,
  pages = {265–270}
}

@article{HensenEtAl_BellTest1,
  title = {Loophole-free Bell inequality violation using electron spins separated by 1.3 kilometres},
  volume = {526},
  ISSN = {1476-4687},
  url = {http://dx.doi.org/10.1038/nature15759},
  DOI = {10.1038/nature15759},
  number = {7575},
  journal = {Nature},
  publisher = {Springer Science and Business Media LLC},
  author = {Hensen,  B. and Bernien,  H. and Dréau,  A. E. and Reiserer,  A. and Kalb,  N. and Blok,  M. S. and Ruitenberg,  J. and Vermeulen,  R. F. L. and Schouten,  R. N. and Abellán,  C. and Amaya,  W. and Pruneri,  V. and Mitchell,  M. W. and Markham,  M. and Twitchen,  D. J. and Elkouss,  D. and Wehner,  S. and Taminiau,  T. H. and Hanson,  R.},
  year = {2015},
  month = oct,
  pages = {682–686}
}

@article{GiustinaEtAl_BellTest2,
  title = {Significant-Loophole-Free Test of Bell’s Theorem with Entangled Photons},
  volume = {115},
  ISSN = {1079-7114},
  url = {http://dx.doi.org/10.1103/PhysRevLett.115.250401},
  DOI = {10.1103/physrevlett.115.250401},
  number = {25},
  journal = {Physical Review Letters},
  publisher = {American Physical Society (APS)},
  author = {Giustina, Marissa and Versteegh, Marijn A.M. and Wengerowsky, S\"{o}ren and Handsteiner, Johannes and Hochrainer, Armin and Phelan, Kevin and Steinlechner, Fabian and Kofler, Johannes and Larsson, Jan-{\AA}ke and Abell\'{a}n, Carlos and Amaya, Waldimar and Pruneri, Valerio and Mitchell, Morgan W. and Beyer, J\"{o}rn and Gerrits, Thomas and Lita, Adriana E. and Shalm, Lynden K. and Nam, Sae Woo and Scheidl, Thomas and Ursin, Rupert and Wittmann, Bernhard and Zeilinger, Anton},
  year = {2015},
  month = dec 
}

@article{ShalmEtAl_BellTest3,
  title = {Strong Loophole-Free Test of Local Realism},
  volume = {115},
  ISSN = {1079-7114},
  url = {http://dx.doi.org/10.1103/PhysRevLett.115.250402},
  DOI = {10.1103/physrevlett.115.250402},
  number = {25},
  journal = {Physical Review Letters},
  publisher = {American Physical Society (APS)},
  author = {Shalm,  Lynden K. and Meyer-Scott,  Evan and Christensen,  Bradley G. and Bierhorst,  Peter and Wayne,  Michael A. and Stevens,  Martin J. and Gerrits,  Thomas and Glancy,  Scott and Hamel,  Deny R. and Allman,  Michael S. and Coakley,  Kevin J. and Dyer,  Shellee D. and Hodge,  Carson and Lita,  Adriana E. and Verma,  Varun B. and Lambrocco,  Camilla and Tortorici,  Edward and Migdall,  Alan L. and Zhang,  Yanbao and Kumor,  Daniel R. and Farr,  William H. and Marsili,  Francesco and Shaw,  Matthew D. and Stern,  Jeffrey A. and Abellán,  Carlos and Amaya,  Waldimar and Pruneri,  Valerio and Jennewein,  Thomas and Mitchell,  Morgan W. and Kwiat,  Paul G. and Bienfang,  Joshua C. and Mirin,  Richard P. and Knill,  Emanuel and Nam,  Sae Woo},
  year = {2015},
  month = dec 
}

@article{GRW,
  title = {Unified dynamics for microscopic and macroscopic systems},
  volume = {34},
  ISSN = {0556-2821},
  url = {http://dx.doi.org/10.1103/PhysRevD.34.470},
  DOI = {10.1103/physrevd.34.470},
  number = {2},
  journal = {Physical Review D},
  publisher = {American Physical Society (APS)},
  author = {Ghirardi,  G. C. and Rimini,  A. and Weber,  T.},
  year = {1986},
  month = jul,
  pages = {470–491}
}

@article{CSL,
  title = {Markov processes in Hilbert space and continuous spontaneous localization of systems of identical particles},
  volume = {42},
  ISSN = {1094-1622},
  url = {http://dx.doi.org/10.1103/PhysRevA.42.78},
  DOI = {10.1103/physreva.42.78},
  number = {1},
  journal = {Physical Review A},
  publisher = {American Physical Society (APS)},
  author = {Ghirardi,  Gian Carlo and Pearle,  Philip and Rimini,  Alberto},
  year = {1990},
  month = jul,
  pages = {78–89}
}

@article{DP,
  title = {Models for universal reduction of macroscopic quantum fluctuations},
  volume = {40},
  ISSN = {0556-2791},
  url = {http://dx.doi.org/10.1103/PhysRevA.40.1165},
  DOI = {10.1103/physreva.40.1165},
  number = {3},
  journal = {Physical Review A},
  publisher = {American Physical Society (APS)},
  author = {Di\'{o}si,  L.},
  year = {1989},
  month = aug,
  pages = {1165–1174}
}

@article{Schlosshauer2005,
  title = {Decoherence,  the measurement problem,  and interpretations of quantum mechanics},
  volume = {76},
  ISSN = {1539-0756},
  url = {http://dx.doi.org/10.1103/RevModPhys.76.1267},
  DOI = {10.1103/revmodphys.76.1267},
  number = {4},
  journal = {Reviews of Modern Physics},
  publisher = {American Physical Society (APS)},
  author = {Schlosshauer,  Maximilian},
  year = {2005},
  month = feb,
  pages = {1267–1305}
}

@article{Joos1985,
  title = {The emergence of classical properties through interaction with the environment},
  volume = {59},
  ISSN = {1434-6036},
  url = {http://dx.doi.org/10.1007/BF01725541},
  DOI = {10.1007/bf01725541},
  number = {2},
  journal = {Zeitschrift f\"{u}r Physik B Condensed Matter},
  publisher = {Springer Science and Business Media LLC},
  author = {Joos,  E. and Zeh,  H. D.},
  year = {1985},
  month = jun,
  pages = {223–243}
}

@article{einselection,
  title = {Decoherence,  einselection,  and the quantum origins of the classical},
  volume = {75},
  ISSN = {1539-0756},
  url = {http://dx.doi.org/10.1103/RevModPhys.75.715},
  DOI = {10.1103/revmodphys.75.715},
  number = {3},
  journal = {Reviews of Modern Physics},
  publisher = {American Physical Society (APS)},
  author = {Zurek,  Wojciech Hubert},
  year = {2003},
  month = may,
  pages = {715–775}
}

@article{jumps,
  title = {Quantum origin of quantum jumps: Breaking of unitary symmetry induced by information transfer in the transition from quantum to classical},
  volume = {76},
  ISSN = {1094-1622},
  url = {http://dx.doi.org/10.1103/PhysRevA.76.052110},
  DOI = {10.1103/physreva.76.052110},
  number = {5},
  journal = {Physical Review A},
  publisher = {American Physical Society (APS)},
  author = {Zurek,  Wojciech Hubert},
  year = {2007},
  month = nov 
}

@article{Darwinism,
  title = {Quantum Darwinism},
  volume = {5},
  ISSN = {1745-2481},
  url = {http://dx.doi.org/10.1038/nphys1202},
  DOI = {10.1038/nphys1202},
  number = {3},
  journal = {Nature Physics},
  publisher = {Springer Science and Business Media LLC},
  author = {Zurek,  Wojciech Hubert},
  year = {2009},
  month = mar,
  pages = {181–188}
}

@article{eavesdropping,
  title = {Eavesdropping on the Decohering Environment: Quantum Darwinism,  Amplification,  and the Origin of Objective Classical Reality},
  volume = {128},
  ISSN = {1079-7114},
  url = {http://dx.doi.org/10.1103/PhysRevLett.128.010401},
  DOI = {10.1103/physrevlett.128.010401},
  number = {1},
  journal = {Physical Review Letters},
  publisher = {American Physical Society (APS)},
  author = {Touil,  Akram and Yan,  Bin and Girolami,  Davide and Deffner,  Sebastian and Zurek,  Wojciech Hubert},
  year = {2022},
  month = jan 
}

@incollection{Valentini_2025,
    author = {Valentini, Antony},
    isbn = {9780198901853},
    title = {The trouble with pilot-wave theory: a critical evaluation},
    booktitle = {Guiding Waves in Quantum Mechanics: One Hundred Years of de Broglie-Bohm Pilot-Wave Theory},
    publisher = {Oxford University Press},
    year = {2025},
    month = {04},
    doi = {10.1093/oso/9780198901853.003.0003},
    url = {https://doi.org/10.1093/oso/9780198901853.003.0003},
}

@article{Bell_1966,
  title = {On the Problem of Hidden Variables in Quantum Mechanics},
  volume = {38},
  ISSN = {0034-6861},
  url = {http://dx.doi.org/10.1103/RevModPhys.38.447},
  DOI = {10.1103/revmodphys.38.447},
  number = {3},
  journal = {Reviews of Modern Physics},
  publisher = {American Physical Society (APS)},
  author = {Bell, John S.},
  year = {1966},
  month = jul,
  pages = {447–452}
}

@inproceedings{deBroglie_1928, 
  title = {La nouvelle dynamique des quanta},
  booktitle = {\'Electrons et Photons: Rapports et Discussions du Cinqui\'eme Conseil de Physique},
  year = {1928},
  author = {de Broglie, L.},
  address = {Brussels, Belgium},
  doi = {10.1017/CBO9781139194983},
}

@article{Bush2024,
  title = {Perspectives on pilot-wave hydrodynamics},
  volume = {125},
  ISSN = {1077-3118},
  url = {http://dx.doi.org/10.1063/5.0210055},
  DOI = {10.1063/5.0210055},
  number = {3},
  journal = {Applied Physics Letters},
  publisher = {AIP Publishing},
  author = {Bush,  John W. M. and Frumkin,  Valeri and Sáenz,  Pedro J.},
  year = {2024},
  month = jul 
}

@inproceedings{ThermoAI,
  title = {Thermodynamic AI and the Fluctuation Frontier},
  url = {http://dx.doi.org/10.1109/ICRC60800.2023.10386858},
  DOI = {10.1109/icrc60800.2023.10386858},
  booktitle = {2023 IEEE International Conference on Rebooting Computing (ICRC)},
  publisher = {IEEE},
  author = {Coles,  Patrick J. and Szczepanski,  Collin and Melanson,  Denis and Donatella,  Kaelan and Martinez,  Antonio J. and Sbahi,  Faris},
  year = {2023},
  month = dec,
  pages = {1–10}
}

@article{Chou2019,
  title = {Analog Coupled Oscillator Based Weighted Ising Machine},
  volume = {9},
  ISSN = {2045-2322},
  url = {http://dx.doi.org/10.1038/s41598-019-49699-5},
  DOI = {10.1038/s41598-019-49699-5},
  number = {1},
  journal = {Scientific Reports},
  publisher = {Springer Science and Business Media LLC},
  author = {Chou,  Jeffrey and Bramhavar,  Suraj and Ghosh,  Siddhartha and Herzog,  William},
  year = {2019},
  month = oct 
}

@inproceedings{Kish2003,
  title = {Quantum computing with analog circuits: Hilbert space computing},
  volume = {5055},
  ISSN = {0277-786X},
  url = {http://dx.doi.org/10.1117/12.497438},
  DOI = {10.1117/12.497438},
  booktitle = {Smart Structures and Materials 2003: Smart Electronics,  MEMS,  BioMEMS,  and Nanotechnology},
  publisher = {SPIE},
  author = {Kish,  Laszlo B.},
  editor = {Varadan,  Vijay K. and Kish,  Laszlo B.},
  year = {2003},
  month = jul,
  pages = {57}
}

@article{Weinberg1989,
  title = {Testing quantum mechanics},
  volume = {194},
  ISSN = {0003-4916},
  url = {http://dx.doi.org/10.1016/0003-4916(89)90276-5},
  DOI = {10.1016/0003-4916(89)90276-5},
  number = {2},
  journal = {Annals of Physics},
  publisher = {Elsevier BV},
  author = {Weinberg,  Steven},
  year = {1989},
  month = sep,
  pages = {336–386}
}

@article{Gisin1989,
    author = "Gisin, N.",
    title = "{Stochastic quantum dynamics and relativity}",
    journal = "Helv. Phys. Acta",
    volume = "62",
    pages = "363--371",
    year = "1989"
}

@article{Polchinski1991,
  title = {Weinberg’s nonlinear quantum mechanics and the Einstein-Podolsky-Rosen paradox},
  volume = {66},
  ISSN = {0031-9007},
  url = {http://dx.doi.org/10.1103/PhysRevLett.66.397},
  DOI = {10.1103/physrevlett.66.397},
  number = {4},
  journal = {Physical Review Letters},
  publisher = {American Physical Society (APS)},
  author = {Polchinski,  Joseph},
  year = {1991},
  month = jan,
  pages = {397–400}
}

@inproceedings{singleTrajectories,
  series = {QIM},
  title = {Observing Single Quantum Trajectories of a Superconducting Qubit},
  url = {http://dx.doi.org/10.1364/QIM.2014.QW1A.1},
  DOI = {10.1364/qim.2014.qw1a.1},
  booktitle = {Research in Optical Sciences},
  publisher = {OSA},
  author = {Siddiqi,  Irfan and Weber,  S. J. and Murch,  K. W. and Chantasri,  A. and Dressel,  J. and Jordan,  A. N.},
  year = {2014},
  pages = {QW1A.1},
  collection = {QIM}
}

@article{Feynman1982,
  title = {Simulating physics with computers},
  volume = {21},
  ISSN = {1572-9575},
  url = {http://dx.doi.org/10.1007/BF02650179},
  DOI = {10.1007/bf02650179},
  number = {6–7},
  journal = {International Journal of Theoretical Physics},
  publisher = {Springer Science and Business Media LLC},
  author = {Feynman,  Richard P.},
  year = {1982},
  month = jun,
  pages = {467–488}
}

@article{montina08,
  title = {Exponential complexity and ontological theories of quantum mechanics},
  author = {Montina, A.},
  journal = {Phys. Rev. A},
  volume = {77},
  issue = {2},
  pages = {022104},
  numpages = {10},
  year = {2008},
  month = {Feb},
  publisher = {American Physical Society},
  doi = {10.1103/PhysRevA.77.022104},
  url = {https://link.aps.org/doi/10.1103/PhysRevA.77.022104}
}

@article{montina11,
  title = {State-space dimensionality in short-memory hidden-variable theories},
  author = {Montina, Alberto},
  journal = {Phys. Rev. A},
  volume = {83},
  issue = {3},
  pages = {032107},
  numpages = {11},
  year = {2011},
  month = {Mar},
  publisher = {American Physical Society},
  doi = {10.1103/PhysRevA.83.032107},
  url = {https://link.aps.org/doi/10.1103/PhysRevA.83.032107}
}

@article{Bielinska2025,
  author  = {Bieli{\'n}ska, Marta Emilia and Eckstein, Micha{\l} and Horodecki, Pawe{\l}},
  title   = {Superluminal signalling and chaos in nonlinear quantum dynamics},
  journal = {New J. Phys.},
  volume  = {27},
  number  = {5},
  pages   = {053002},
  year    = {2025},
  doi     = {10.1088/1367-2630/adcae1}
}

@article{Das2022,
  author  = {Das, Swetamber and Green, Jason R.},
  title   = {Density matrix formulation of dynamical systems},
  journal = {Phys. Rev. E},
  volume  = {106},
  number  = {5},
  pages   = {054135},
  year    = {2022},
  doi     = {10.1103/PhysRevE.106.054135}
}

@article{Das2023,
  author  = {Das, Swetamber and Green, Jason R.},
  title   = {Speed limits on deterministic chaos and dissipation},
  journal = {Phys. Rev. Res.},
  volume  = {5},
  number  = {1},
  pages   = {L012016},
  year    = {2023},
  doi     = {10.1103/PhysRevResearch.5.L012016}
}

@article{Skinner2019,
  author  = {Skinner, Brian and Ruhman, Jonathan and Nahum, Adam},
  title   = {Measurement-Induced Phase Transitions in the Dynamics of Entanglement},
  journal = {Phys. Rev. X},
  volume  = {9},
  number  = {3},
  pages   = {031009},
  year    = {2019},
  doi     = {10.1103/PhysRevX.9.031009}
}

@article{Noel2022,
  author  = {Noel, Crystal and Niroula, Pradeep and Zhu, Daiwei and Risinger, Andrew and Egan, Laird and Biswas, Debopriyo and Cetina, Marko and Gorshkov, Alexey V. and Gullans, Michael J. and Huse, David A. and Monroe, Christopher},
  title   = {Measurement-induced quantum phases realized in a trapped-ion quantum computer},
  journal = {Nat. Phys.},
  volume  = {18},
  pages   = {760--764},
  year    = {2022},
  doi     = {10.1038/s41567-022-01619-7}
}

@article{Anza2021,
  author  = {Anz{\`a}, Fabio and Crutchfield, James P.},
  title   = {Beyond density matrices: Geometric quantum states},
  journal = {Phys. Rev. A},
  volume  = {103},
  number  = {6},
  pages   = {062218},
  year    = {2021},
  doi     = {10.1103/PhysRevA.103.062218}
}

@article{BohmBub1966,
  title = {A Proposed Solution of the Measurement Problem in Quantum Mechanics by a Hidden Variable Theory},
  volume = {38},
  url = {https://doi.org/10.1103/RevModPhys.38.453},
  DOI = {10.1103/RevModPhys.38.453},
  number = {3},
  journal = {Reviews of Modern Physics},
  publisher = {American Physical Society (APS)},
  author = {Bohm, David and Bub, Jeffrey},
  year = {1966},
  month = jul,
  pages = {453--469}
}

@article{Pearle1976,
  title = {Reduction of the state vector by a nonlinear {Schr\"odinger} equation},
  volume = {13},
  url = {https://doi.org/10.1103/PhysRevD.13.857},
  DOI = {10.1103/PhysRevD.13.857},
  number = {4},
  journal = {Physical Review D},
  publisher = {American Physical Society (APS)},
  author = {Pearle, Philip},
  year = {1976},
  month = feb,
  pages = {857--868}
}

@article{WoottersZurek1982,
  author  = {Wootters, W. K. and Zurek, W. H.},
  title   = {A single quantum cannot be cloned},
  journal = {Nature},
  volume  = {299},
  pages   = {802--803},
  year    = {1982},
  doi     = {10.1038/299802a0}
}

@article{Dieks1982,
  author  = {Dieks, D.},
  title   = {Communication by {EPR} devices},
  journal = {Phys. Lett. A},
  volume  = {92},
  number  = {6},
  pages   = {271--272},
  year    = {1982},
  doi     = {10.1016/0375-9601(82)90084-6}
}

@book{NielsenChuang,
  author    = {Nielsen, Michael A. and Chuang, Isaac L.},
  title     = {Quantum Computation and Quantum Information},
  edition   = {10th Anniversary},
  publisher = {Cambridge University Press},
  address   = {Cambridge},
  year      = {2010}
}

@article{Gisin1990,
  author  = {Gisin, N.},
  title   = {Weinberg's non-linear quantum mechanics and supraluminal communications},
  journal = {Phys. Lett. A},
  volume  = {143},
  number  = {1-2},
  pages   = {1--2},
  year    = {1990},
  doi     = {10.1016/0375-9601(90)90786-N}
}

@article{Valentini2002,
  author  = {Valentini, Antony},
  title   = {Signal-locality in hidden-variables theories},
  journal = {Phys. Lett. A},
  volume  = {297},
  number  = {5-6},
  pages   = {273--278},
  year    = {2002},
  note    = {arXiv:quant-ph/0106098}
}

@article{Touil2024,
  author  = {Touil, Akram and Anza, Fabio and Deffner, Sebastian and Crutchfield, James P.},
  title   = {Branching States as The Emergent Structure of a Quantum Universe},
  journal = {Quantum},
  volume  = {8},
  pages   = {1494},
  year    = {2024},
  doi     = {10.22331/q-2024-10-10-1494}
}
